\documentclass[preprint2]{proto}
\usepackage{times}
\usepackage{epsfig}
\newcommand{\refs}{\par\noindent\hangindent=1pc\hangafter=1}
\begin{document}

\title{\textbf{\LARGE Stellar Properties of Embedded Protostars}}

\author {\textbf{\large R. J. White}}
\affil{\small\em University of Alabama in Huntsville}
\author {\textbf{\large T. P. Greene}}
\affil{\small\em National Aeronautics and Space Administration at Ames Research Center}
\author {\textbf{\large G. W. Doppmann}}
\affil{\small\em Gemini Observatory}
\author {\textbf{\large K. R. Covey}}
\affil{\small\em University of Washington}
\author {\textbf{\large L. A. Hillenbrand}}
\affil{\small\em California Institute of Technology}

\begin{abstract}
\baselineskip = 11pt
\leftskip = 0.65in
\rightskip = 0.65in
\parindent=1pc

{\small Protostars are precursors to the nearly fully assembled T-Tauri and 
  Herbig  
Ae/Be type stars undergoing quasi-static contraction towards the zero-age
main sequence; they are in the process of acquiring the majority of their
stellar mass.  Although numerous young stars with spatially extended
envelope-like structures appear to fit this description, their high
extinction has inhibited observers from directly measuring their stellar
and accretion properties and confirming that they are in fact in the main
phase of mass accretion (i.e., true protostars).  Recently, however, high
dispersion spectrographs on large aperture telescopes have allowed
observers to begin studying the stellar and accretion properties of 
a subset of these stars, commonly referred to as Class I stars.  
In this Chapter, we summarize the newly determined properties of Class I
stars and compare them with observations of Class II stars, which are 
the more optically revealed T Tauri stars, to better understand the
relative evolutionary state of the two classes.  Class I stars have 
distributions of spectral types and stellar luminosities that are similar
to those of Class II stars, suggesting similar masses and ages.  The
stellar luminosity and resulting age estimates, however, are especially
uncertain given the difficulty in accounting for the large extinctions,
scattered light emission and continuum excesses typical of Class I
stars.  Several candidate Class I brown 
dwarfs are identified.  Class I stars appear to rotate somewhat more
rapidly than T Tauri stars, by roughly a factor of 2 in the mean.
Likewise, the disk accretion rates inferred from optical excesses and
Br$\gamma$ luminosities are similar to,  
but larger in the mean by a factor of a few than, the disk accretion rates of 
T Tauri stars.  There is some evidence that the disk accretion rates of Class 
I stars are more distinct from T Tauri stars within the $\rho$ Ophiuchi star 
forming region than in others (e.g., Taurus-Auriga), suggesting a possible 
environmental influence.  The determined disk accretion rates are nevertheless
1-2 orders of magnitude less than the mass infall rates predicted by
envelope models.
In at least a few cases the discrepancy appears to be caused by T Tauri stars 
being misclassified as Class I stars because of their edge-on disk orientation.
In cases where the envelope density and infall velocity have been determined
directly and unambiguously, the discrepancy suggests that the stellar mass is
not acquired in a steady-state fashion, but instead through brief outbursts 
of enhanced accretion.  If the ages of some Class I stars are in fact as
old as T Tauri stars, replenishment may be necessary to sustain the 
long-lived envelopes, possibly via continued dynamical interactions with
cloud material. 
 \\~\\~\\~}
\end{abstract}  

{\centering
\section{\textbf{THE DISCOVERY AND CLASSIFICATION OF PROTOSTARS}}
}

The early phases of star and planet formation are difficult to observe
because this process occurs while the protostar is buried within its
natal molecular cloud material.  Nevertheless, infrared and
submillimeter observations, which are able to penetrate this
high extinction material, have revealed much about the bolometric 
luminosities, spectral energy distributions (SEDs), and circumstellar
material of embedded 
young stars (e.g., {\it Lada and Wilking}, 1984; {\it Myers et al.}, 1987;
{\it Wilking et al.}, 1989; {\it Kenyon et al.}, 1990;
{\it Andr\'{e} and Montmerle}, 1994; {\it Motte and Andr\'{e}}, 2001; 
{\it Onishi et al.}, 2002; {\it Andrews and Williams}, 2005).  The earliest
of these observations spurred development 
of the theory of isolated low mass star formation, advancing initial 
considerations of the collapse of a singular isothermal sphere (e.g., 
{\it Shu}, 1977) to include circumstellar disks and envelopes ({\it Cassen 
and Moosman}, 1981; {\it Terebey et al.}, 1984; {\it Adams et al.}, 1987).

An easy marriage of observation and theory was found by equating
different stages of this theoretical evolutionary process with
observed differences in the spectral energy distributions of very
young stars.  Four classes have been proposed (Class 0, I, II, and
III), and are now commonly used to classify young stars.
In this proposed scheme, Class 0 stars are cloud cores that are just
beginning their protostellar collapse, Class I stars are embedded within an 
``envelope'' of circumstellar material, which is infalling, accumulating in
a disk, and being channeled onto the star, Class II stars are nearly fully
assembled stars undergoing pure disk accretion with perhaps some
evidence for tenuous amounts of envelope material and, finally, Class III
stars are post-accretion but still pre-main sequence stars.  The
Class II and Class III stars are also known as classical T Tauri stars 
and weak-lined T Tauri stars, respectively.  It is believed that the
majority of the stellar mass is acquired prior to the Class II phase;
these younger stars are thus considered to be the true
``protostars.''

Despite the discretization of the Class classification scheme,
there is a continuum of circumstellar evolutionary states and thus a
continuum of observational properties exhibited by young stars.
Fig. 1 illustrates two popular criteria used to segregate the
Classes, bolometric temperature ($T_{bol}$, defined as the temperature
of a blackbody with the same mean frequency as the observed SED; {\it Myers
  and Ladd}, 1993) and infrared spectral slope ($\alpha = d$log[$\lambda
F_{\lambda}$] / [$d$log$\lambda$], typically determined over the wavelength
interval 2 to 25 $\mu$m; {\it Lada and Wilking}, 1984; {\it Lada}, 1987),
plotted against one another.  Class I stars are distinguished from Class II
stars as having $\alpha > 0.0$ or $T_{bol} <650~K$; their SEDs rise into
the infrared.  A subsample of ``flat spectrum" or
``transitional Class I/II'' stars are often distinguished as those
with $-0.3 < \alpha < 0.3$ or $650 < T_{bol} < 1000~K$.  However,
since these criteria are based on observations which
typically do not spatially resolve the circumstellar structures, it is
not clear that the observed SED differences truly correspond to
distinct evolutionary stages.  Line of sight orientation or unresolved
companions, as examples, can significantly alter the observed SED. 


Studies of the emergent SEDs at wavelengths $\ga 10 \mu$m have provided 
important, albeit ambiguous, constraints on the circumstellar dust 
distributions for Class I stars.  Although existing data are based on 
relatively low spatial resolution observations from IRAS and ISO, with 
the promise of the Spitzer Space Telescope ({\it Werner et al.}, 2004) 
currently being realized, a single generic representation of Class I and 
some I/II stars has been developed.  Models incorporating infalling, 
rotating envelopes with mass infall rates on the order of $10^{-6}$
M$_{\odot}$/yr predict SEDs that are consistent with observations
({\it Adams et al.}, 1987; {\it Kenyon et al.}, 1993a; {\it Whitney et
  al.}, 1997, 2003).
However, only in a few cases are these mass infall rates supported by 
kinematic measurements of spatially resolved envelope structures
(e.g., {\it Gregerson et al.}, 1997).  For some young stars 
whose SEDs can be explained by spherically-symmetric dust distributions, 
it has been suggested that nearly edge-on flared disk models may also be 
able to reproduce the SEDs (e.g., {\it Chiang and Goldreich}, 1999; {\it
Hogerheijde and Sandell}, 2000).  


\epsfig{file=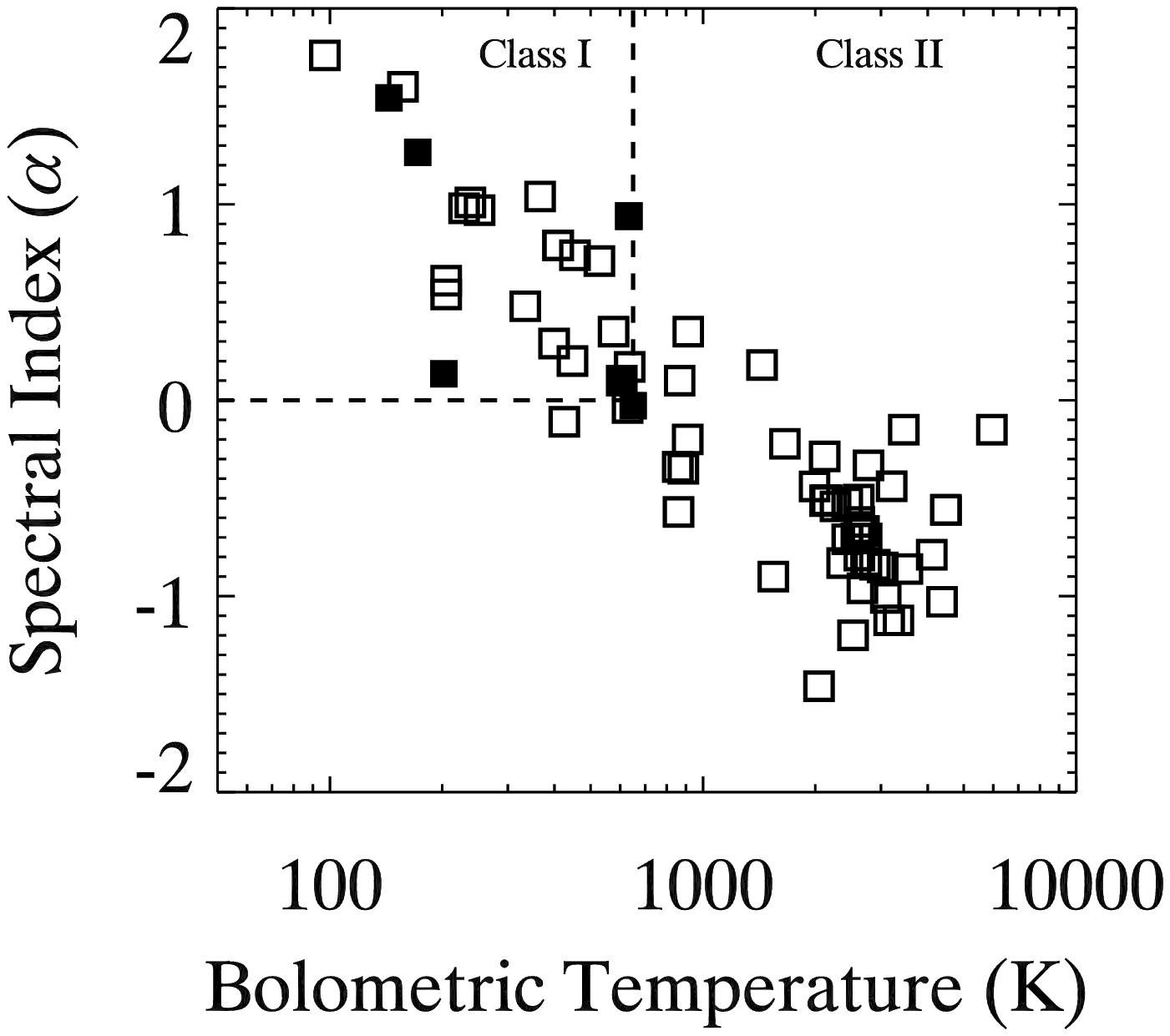, width = 3.1in}
\noindent
{\small Fig. 1. Spectral index versus bolometric temperature for young
 stars observed spectroscopically ({\it White and Hillenbrand}, 2004; {\it
 Doppmann et al.}, 2005) in Taurus and $\rho$ Ophiuchi and which have
 both evolutionary diagnostics determined.  Open symbols
 correspond to stars with detected photospheric features from which stellar
 properties can be extracted; filled symbols are too heavily veiled to
 extract these features.  Class I stars are bolometrically cold ($T_{bol} 
 < 650$ K) and have rising mid-infrared energy distributions ($\alpha > 
 0.0$).  The
 new spectroscopic observations extend well into the Class I regime.}
\bigskip

Whether the observable diagnostics trace distinct evolutionary
states bears directly on the issue of whether Class I stars are in fact
younger than Class II stars, as is often assumed, or whether they are
simply less environmentally developed; they could be T Tauri age stars
still (or perhaps just currently) embedded within circumstellar material.  
What is needed is an understanding of the {\it stellar} properties of 
these systems.  To date, stellar properties such as mass and age have 
been derived for Class I stars predicated on the assumption that they 
are in the main stage of infall (see e.g., {\it Evans}, 1999), that this
material is accumulating in a circumstellar disk, and then accretes onto
the star at a rate sufficient to match the bolometric 
luminosity (defined as the luminosity of a star's entire energy 
distribution).  Given an assumed mass, the age of the star is then 
simply the mass divided by the mass infall rate (0.6 M$_\odot$ / 
$3\times 10^{-6}$ M$_\odot$/yr $= 2\times 10^{5}$ yr).  
Buttressing the argument for the extreme youth of Class I stars is the 
relative number of Class I, II, and III stars in clouds such as Tau-Aur.  
As discussed by {\it 
Benson and Myers} (1989) and {\it Kenyon et al.} (1990), the relative ages 
of stars in different stages can be inferred from their relative numbers,
assuming a constant star formation rate.  For 
Taurus-Auriga, there are 10 times fewer Class I stars than Class II and 
Class III stars, implying the Class I phase must be 10 times shorter, 
leading to age estimates of $\sim 2\times 10^5$ yr assuming typical ages 
of $\sim 2\times 10^6$ yr for the Class II/III population, as inferred from 
the Hertzsprung-Russell diagram (e.g., {\it Kenyon and Hartmann}, 1995).

A more robust confirmation that Class I stars are bona-fide protostars 
would be an unambiguous demonstration that they are acquiring mass at a
much higher rate than Class II stars.  Although the mass infall rates
inferred (indirectly, in most cases) for Class I stars are roughly 2 
orders of magnitude larger than the disk accretion rates determined for 
Class II stars ($\sim 10^{-8}$ M$_\odot$/yr; e.g., {\it Gullbring et al.}, 
1998), it has not yet been shown that the infalling material is 
channeled through the disk and onto the star at this
same prodigious rate.  Under the assumption that these two 
rates are the same leads to a historical difficulty with the
Class I paradigm - the so-called ``luminosity
problem.''  As first pointed out by {\em Kenyon et al. (1990)}, if the
material infalling from the envelope is channeled through the disk via
steady-state accretion and onto the star, the accretion luminosity
would be dominant at roughly 10
times the luminosity emitted from the photosphere.  However, Class I
stars, at least in Tau-Aur, do not have integrated luminosities 
substantially different from those of neighboring T Tauri stars.  
Several reconciliations have been proposed, including disk accretion 
which is not steady-state, very low mass (i.e., substellar) central 
masses, or simply erroneously large mass infall rates.  Direct 
measurement, rather than indirect inference, of both the stellar and 
the accretion luminosities of Class I stars is needed to distinguish
between these.

The most straightforward way to unambiguously determine the stellar 
and accretion properties of young stars at any age is to observe their
spectra at wavelengths shorter than $\sim 3 \mu$m where the peak flux from
the stellar photosphere is emitted.  
While this has been possible for over five decades for Class II
stars, the faintness of Class I stars at optical and near-infrared
wavelengths have made it difficult to obtain high resolution, high
signal-to-noise observations necessary for such measurements.  The
development of sensitive spectrographs mounted on moderate to large
aperture telescopes now allow direct observations of Class I and I/II
photospheres via light scattered through circumstellar envelopes.
These observational windows provide an 
opportunity to study Class I stars with the same tools and techniques
developed for the study of Class II stars.
\bigskip

{
\centering
\section{\textbf{PHOTOSPHERES AND ACCRETION}}
}

Detailed spectroscopic studies of young stars much less embedded than
protostars (e.g., T Tauri stars) have provided much of the
observational basis for theories of how stars are assembled and how
they interact with their environment.  The spectrum of the canonical 
young, accreting, low-mass star consists of a late-type photosphere with
strong emission-lines and excess continuum emission (i.e., veiling) at 
optical and infrared wavelengths.  At optical
wavelengths, measurement of this excess emission, which is attributed 
to high temperature regions generated in the accretion flow, 
provides a direct
estimate of the mass accretion rate and constrains physical conditions
of accretion shock models (see the chapter by {\it Bouvier et al.}).  At
infrared wavelengths, measurement of the excess thermal emission from
warm circumstellar dust reveals structural information of the inner
accretion disk ({\it Najita}, 2004; {\it Muzerolle et al.}, 2004; {\it 
Johns-Krull et al.}, 2003).  Additionally, the strengths and profile 
shapes of permitted emission-lines delimit how circumstellar material 
is channeled onto the stellar surface ({\it Calvet and Hartmann}, 1992;
{\it Muzerolle et al.}, 1998, 2001), while density-sensitive
forbidden emission lines trace how and how much mass is lost in
powerful stellar jets (e.g., {\it Hartigan et al.}, 1995).
Perhaps most importantly, extraction of the underlying photospheric 
features permit the determination of precise stellar properties 
(T$_{eff}$, log $g$,
[Fe/H]), which can be compared to evolutionary models to determine 
stellar masses and ages.  Doppler broadening of these features also
provides a measure of the stellar rotation rate ($v$sin$i$), which is
important for tracing the evolution of angular momentum.  
Spectroscopic observations at visible and near-infrared wavelengths 
are 2 powerful tools for studying a young star's photospheric 
properties and its circumstellar accretion, if realizable.

\medskip
\subsection{Visible Light}

Although observations at visible or optical wavelengths ($\lesssim 1\, 
\mu$m) are especially
challenging for highly extincted stars, there are nevertheless two
motivations for pursuing this.  First, visible light is dominated by
emission from both the photosphere and high temperature accretion
shocks; it therefore offers the most direct view of stellar properties
and accretion luminosity.  Second, for small dust grains
($\lesssim 1\, \mu$m), visible light scatters more efficiently than
infrared light.  Thus, even if the direct line-of-sight
extinction is too large for an embedded star to be observed directly,
the cavities commonly seen in the envelopes surrounding Class I stars
(e.g., {\it Padgett et al.}, 1999) may permit observations of the
photosphere and inner accretion processes through scattered light.
This is only feasible in low column density star forming
environments like Taurus-Auriga where the young stars are not deeply
embedded within the large-scale molecular cloud.

Recognition of faint but nevertheless detectable emission from
these embedded stars inspired several low resolution spectroscopic
studies with the aim of putting the first solid constraints on the
stellar and accretion properties of suspected protostars.  
This work began even prior to the
now established Class classification scheme; some of the first
embedded young stars were identified by the strong stellar jets which
they powered (e.g., {\it Cohen and Schwartz}, 1983; {\it Graham}, 1991).  These
stars typically had nearly featureless continua with strong
emission lines superimposed.  {\em Mundt et al. (1985)} obtained an
optical spectra of the Class I star L1551 IRS 5 in Taurus-Auriga and
identified the star as a G or K spectral type (but see {\it Osorio et al},
2003); the emission line features showed P Cygni-like profiles suggestive 
of a strong outflowing wind.  More recently, {\em Kenyon et al. (1998)}
reported spectroscopic observations for 10 of the Class I stars in
Taurus-Auriga, detecting M spectral type features (i.e., TiO bands) in
several and strong emission line features in all.
These initial spectroscopic studies suggested that at least
some Class I stars resemble their more evolved T Tauri star
counterparts (Class II stars), but with heavily veiled spectra and
strong emission lines.  Unfortunately, the limited numbers of stars
with revealed spectroscopic features, due in part to the low
spectral resolution of the observations, precluded accurate determination 
of stellar properties and specific mass accretion and mass outflow rates
for unbiased comparisons with the more optically revealed T Tauri stars.

\setcounter{figure}{1}
\begin{figure*}
 \epsscale{2.0}
 \plotone{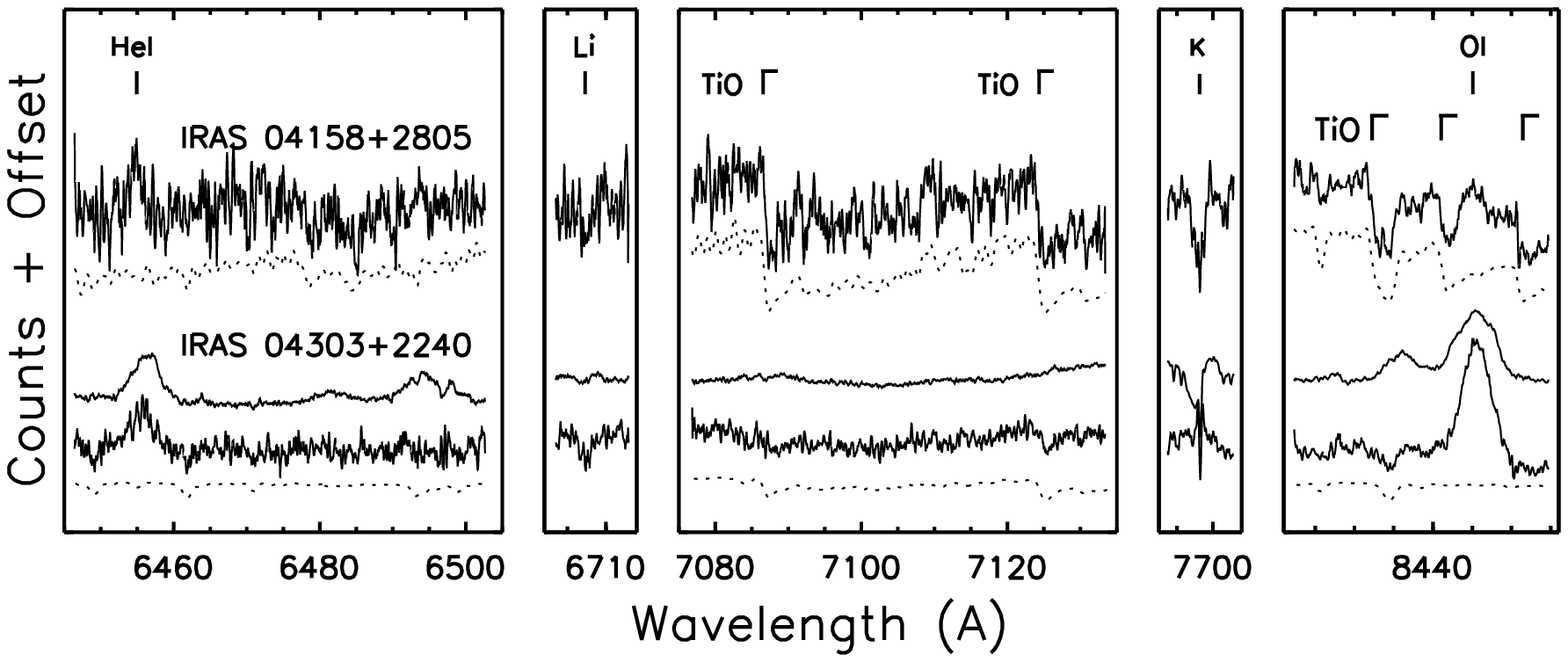}
 \caption{\small Portions of the Keck/HIRES spectra from {\it White
 and Hillenbrand} (2004).  IRAS 04158+2805 ($\alpha = +0.71$) has a very
 cool spectral type ($\sim$ M6) and possibly a substellar mass.  The two
 epochs of IRAS 04303+2240 ($\alpha = -0.35$) show dramatic variations in
 the veiling and the inferred mass accretion rate; the heavily veiled
 spectrum is less noisy because the star was also much brighter.   Spectra
 of the best fit dwarf stars, veiled and rotationally broadened, are shown
 as dotted lines.} 
\end{figure*}

\subsection{Near-Infrared Light}

The development of infrared detector technology during the 1980s and
90s has provided another valuable tool for the study of protostars.  Since
many stars form in high extinction clouds that block nearly all
visible light (e.g., $\rho$ Ophiuchi, Serpens), they are not
amenable to study at visible wavelengths.  It has been recognized for
some time that late-type stellar photospheres exhibit a number of
atomic and molecular features in the 2 -- 2.4 $\mu$m wavelength region
($K$ band) which are diagnostic of effective temperatures and surface
gravities ({\it Kleinmann and Hall}, 1986; {\it Wallace and Hinkle}, 1996),
and can be used to measure stellar projected rotational and radial
velocities.  Interstellar dust is also relatively transparent in this
wavelength region, $A_{K} \simeq 0.1$ $A_{V}$ (in magnitudes), permitting
spectroscopic observations of even highly
extinguished young stars in nearby dark clouds to be obtained.  However,
the near-infrared spectra of embedded young stars 
are frequently complicated by the presence of thermal emission from
warm dust grains in their inner circumstellar disks or inner envelope
regions.  This excess circumstellar emission can be several times
greater than the photospheric flux of an embedded young star in the
$K$ band wavelength region, causing an increased continuum level that
veils photospheric features.

Initial near-infrared observations at low resolution found that the
CO absorption features at $2.3 \mu$m could be identified less
often for Class I stars than Class II stars ({\it Casali and Matthews},
1992; {\it Casali and Eiroa}, 1996).  This was interpreted as Class I stars
having larger 
near-infrared excess emission than Class II stars, possibly because of 
more luminous circumstellar disks caused by larger mass accretion rates 
or alternatively, envelope emission ({\it Greene and Lada}, 1996; 
{\it Calvet et al.}, 1997).  {\em Muzerolle et al.} (1998)
demonstrated that the Br$\gamma$ (2.166 $\mu$m) luminosity correlates well
with the total accretion luminosity, and used this relation to measure the
the first accretion luminosities for Class I stars.  The determined accretion 
luminosities were only a small fraction ($\sim$ one-tenth) of the bolometric
luminosity; assuming a typical T Tauri star mass and a radius, these 
accretion luminosities correspond to mass accretion rates that are overall
similar to those of T Tauri stars ($\sim 10^{-8}$ M$_\odot$/yr).  With 
regard to stellar features, {\it Greene and Lada} (1996) and {\it Luhman
  and Rieke} (1999) showed that at least $\sim 25$\% of Class I and 
flat-spectrum stars exhibited temperature sensitive photospheric
features, suggesting that stellar properties could potentially be 
determined directly (see also Ishii et al., 2004).
As with early optical observations, however, low spectral resolution and
large infrared excesses prevented accurate extraction of these
properties.  More recently, {\it Nisini et al.} (2005) presented spectra of
3 Class I stars in R CrA at moderate resolution ($R \sim 9000$),
sufficient to measure the amount of continuum excess and assign spectral
types (i.e., temperatures), but (in this case) insufficient to measure 
radial and rotational velocities.

\subsection{\textbf{The Promise of High Resolution Spectra}}

Fortunately, high dispersion spectrographs on large aperture
telescopes have allowed observers to begin studying the stellar and
accretion properties of embedded low mass protostars in detail, at
both optical and near-infrared wavelengths.  Initial measurements
demonstrated that the key to spectroscopically resolving faint
photospheric features, given the large continuum excess emission, is
high signal-to-noise, high dispersion spectroscopy ({\it Greene and 
Lada}, 1997, 2000, 2002; 
{\it Doppmann et al.}, 2003).  This pioneering work showed that
fundamental photospheric diagnostics (temperatures, surface gravities,
rotational velocities) and circumstellar features (continuum excesses,
emission line luminosities) could be measured nearly as precisely for
Class I stars as for Class II stars.  The small number of Class I
stars ``revealed'' however, inhibited statistically meaningful
comparisons with T Tauri stars to search for evolutionary differences.

Very recently the situation changed dramatically with two large
surveys of embedded stars.  {\it White and Hillenbrand} (2004; hereafter 
{\it WH04}) conducted a high resolution ($R \simeq 34,000$) optical 
spectroscopic study of 36 ``environmentally young'' stars in Taurus-Auriga 
(Tau-Aur).  WH04 
classify stars as ``environmentally young'' if they are either Class I 
stars or power a Herbig-Haro flow.  Their 
sample consisted of 15 Class I stars and 21
Class II stars; they detected photospheric features in 11 of the Class
I stars and all of the Class II stars.  Fig. 2 shows three optical
spectra from this survey.
{\it Doppmann et al.} (2005; hereafter {\it D05}) conducted 
a complementary high resolution ($R \simeq 18,000$) $K$ band study of 
52 Class I and flat-SED stars, selected from 5 nearby star forming
regions - Taurus-Auriga (Tau-Aur), $\rho$ Ophiuchi ($\rho$ Oph), 
Serpens, Perseus, and 
R Corona Australi (R CrA).  Forty-one of the 52 stars were found to have
photospheric absorption features from which stellar properties and
excess emission could be measured.  Fig. 3 shows 3 near-infrared
spectra from this survey.

{\centering
\section{\textbf{SPECTROSCOPIC PROPERTIES OF PROTOSTARS REVEALED}}
}

In this section, we present a combined assessment of the stellar and
accretion properties of Class I and transitional Class I/II stars as
inferred in the {\it WH04} and {\it D05} studies, including other results 
when applicable.  We note that the two primary studies were able to
determine astrophysical properties for 6 of the same stars, permitting
a direct comparison of the two techniques.  Agreement is good for 5 of
the 6 overlapping stars (in $v$sin$i$ and effective temperature); the
1 discrepancy occurs in a heavily veiled, very low signal-to-noise (optical) 
observation; the infrared properties are adopted in this case.

\begin{figure*}
 \epsscale{2.0}
 \plotone{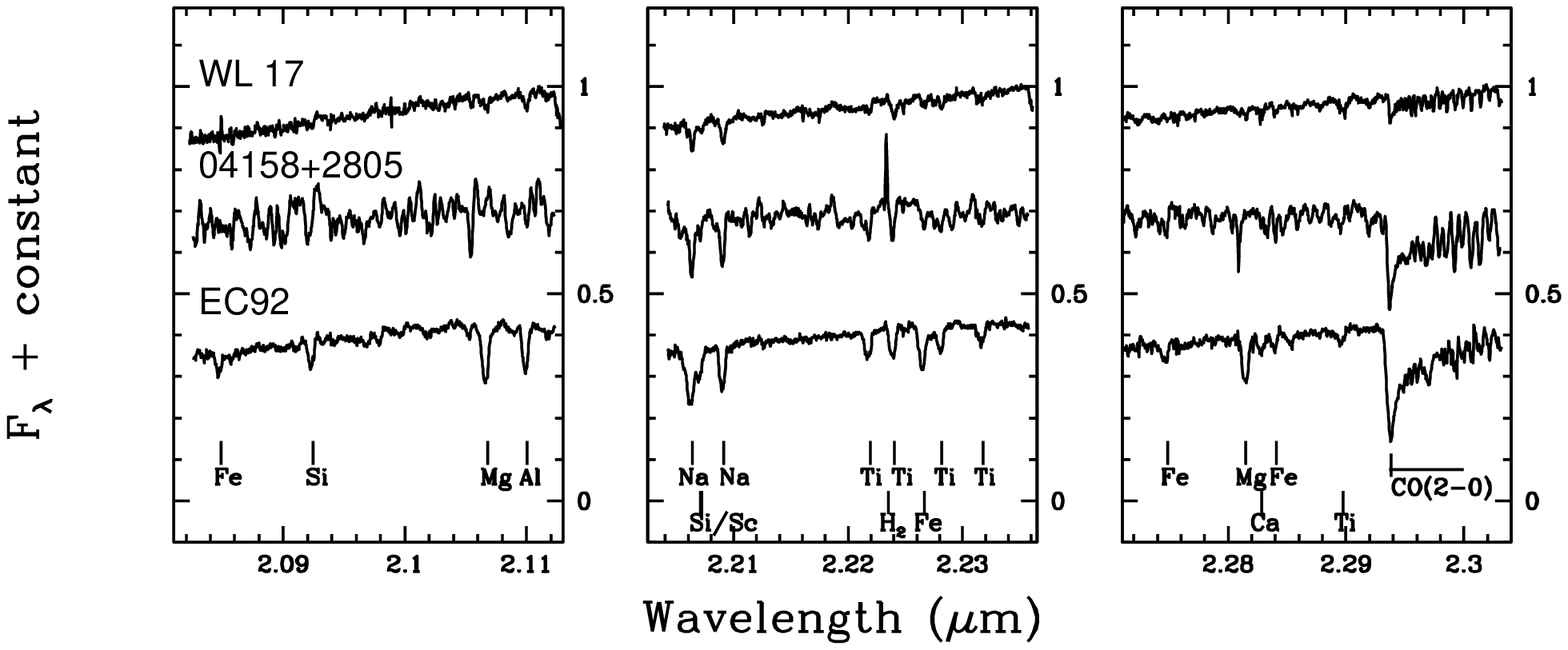}
 \caption{\small High resolution near-infrared spectra of the
   embedded protostars from {\it Doppmann et al.} (2005).  WL 17 ($\alpha =
   +0.42$) is heavily veiled, IRAS 04158+2805 ($\alpha = +0.71$) has a very
   cool spectral type ($\sim$ M6) and possibly a substellar mass, and EW 92
   is moderately rapidly rotating ($v$sin$i$ = 47 km/s).}
\end{figure*}

To identify how stellar and circumstellar properties change as a star
evolves through the proposed evolutionary scheme, we present the inferred
properties as a function of the evolutionary diagnostic $\alpha$, the infrared
spectral index.  We adopt this diagnostic simply because it is 
available for most
of the stars observed.  In addition to the primary samples of {\it WH04} and {\it D05},
we include a sample of accreting Class II stars from Tau-Aur (as assembled
in {\it WH04}) and $\rho$ Ophiuchi (assembled in {\it Greene and Lada}, 1997 and
{\it Doppmann et al.}, 2003), whose properties have
been determined from high dispersion spectra as well.  When available, we
selected values of $\alpha$ calculated from observations at 2 and 25 $\mu$m; 
when such measurements are not available, we use $\alpha$ values calculated 
over a smaller wavelength interval (typically based on ISO observations
extending to 14 $\mu$m).  Specifically, stars in Tau-Aur, NGC 1333 and
R CrA have 2-25 $\mu$m $\alpha$ values determined from IRAS
observations by {\it Kenyon and Hartmann} (1995), {\it Ladd et al.} (1993), 
and {\it Wilking et al.} (1992), respectively.  Serpens and
$\rho$ Oph stars have 2-14 $\mu$m $\alpha$ values from the work of {\em
  Kaas et al} (2004) and {\em Bontemps et al} (2001).  As emphasized in the
introduction, however, all evolutionary diagnostics are subject to
significant biases, which can mask subtle trends.  Thus, we will primarily
make ensemble comparisons between stars classified as Class I stars ($\alpha 
> 0.0$) and stars classified as Class II stars.

\medskip
\subsection{Stellar Masses}

Historically, the masses of embedded young stars have been poorly determined
by observations, since in most cases, the only measurable property was the
bolometric luminosity from the (often poorly determined) SED.  
IRAS surveys of the Tau-Aur, $\rho$ Oph, R CrA, and Chamaeleon I dark
clouds revealed populations of Class I embedded stars in each region
with bolometric luminosities spanning from below 0.1 $L_{\odot}$ to
approximately 50 $L_{\odot}$, with a median value near 1 $L_{\odot}$. 
({\it Kenyon et al.}, 1990; {\it Wilking et al.}, 1989, 1992; {\it Prusti
  et al.}, 1992).  Converting these luminosities to mass estimates requires
 an assumed mass-luminosity relation, which strongly depends upon age, and
 an assumed accretion luminosity.  
If the embedded Class I stars in these regions have
luminosities dominated by accretion, then their masses can be approximated
by applying the spherical accretion luminosity relation $L_{\rm bol} =
L_{\rm acc} = G M_{*} \dot{M} / R_{*}$.  Adopting an infall rate of
2$\times10^{-6}$ M$_\odot$/yr and an protostellar mass--radius relation
(e.g., {\it Adams et al.},  1987; {\it
  Hartmann}, 1998) leads to a mass of 0.5 M$_{\odot}$ (at a radius of 3
R$_\odot$) for a 10 $L_{\odot}$ star.  
Thus, only
the most luminous Class I stars would have inferred masses consistent 
with those of T Tauri stars (0.1 - a few $M_{\odot}$; {\it Kenyon and
  Hartmann}, 1995; {\it Luhman and Rieke}, 1999); the majority would have
masses $\lesssim 0.1$ 
M$_\odot$.  Although there remain considerable 
uncertainties in the calculated bolometric luminosities and the 
prescription for accretion for Class I stars, the emerging census
suggests a "luminosity problem" as described in Section 1; the typical
Class I star is under-luminous relative to what is expected for a canonical 
T Tauri size star (in mass and radius) accreting at the predicted envelope
infall rates.
The luminosity problem is most severe in the Tau-Aur star forming
region ({\it Kenyon et al.}, 1990); there is tentative evidence for a 
regional dependence upon the distribution of bolometric luminosities 
of Class I stars.  One proposed solution to the luminosity problem
is that Class I stars are in fact much lower in mass than T Tauri 
stars (i.e., brown dwarfs), either because they are forming from less 
massive cores/envelopes or because
they have yet accreted only a small fraction of their final mass.  These
possible solutions introduce yet additional problems, however.  If almost
all Class I "stars" are producing brown dwarfs, then "star" formation in 
most regions must have already ceased, implying an unexpected mass 
dependent formation time-scale.  Alternatively, the hypothesis that 
Class I stars have accreted only a small fraction of their final mass is
inconsistent with their relatively low envelope masses ($\sim$ 
$0.1$ M$_\odot$), estimated from millimeter wavelength observations 
(e.g., {\it Motte and Andr\'e}, 2001).  Accurately determined stellar 
mass estimates are needed to test this proposed yet problematic solution 
to the luminosity problem.

One direct way to estimate the mass of a young star is to observationally
determine its stellar effective temperature and luminosity and then 
compare them with the predictions of pre-main sequence
(PMS) evolutionary models.  The recent optical and near-IR 
spectroscopic studies of WH04 and D05
have been able to achieve this for the first time
for several dozen embedded young stars in nearby dark clouds.  Since 
low mass, fully-convective stars primarily evolve in luminosity while young
(e.g., {\it Baraffe et al.}, 1998), temperature is especially important in 
determining a young star's mass.  In most cases, the temperature estimates 
for the Class I stars are as precisely determined as those for T 
Tauri stars ($\sim 150$ K), which translates into similar uncertainties 
in the inferred stellar masses (a few tens of percent), but large 
systematic uncertainties remain (e.g., temperature scale - see the 
chapter by {\it Mathieu et al.}; effects of accretion - {\em Siess et al.}, 
1999; {\em Tout et al.}, 1999).
The uncertainties in the stellar luminosities of embedded stars are, on
the other hand, typically much larger than those for T Tauri stars.  
{\it WH04} estimated stellar luminosities by performing a bolometric correction
from near-infrared $J$-band ($\lambda \simeq 1.25 \mu$m) photometric data, which
is expected to be least contaminated by circumstellar excesses (see,
however, {\it Cieza et al. 2005}); extinctions
were determined by comparing the observed $J-H$ colors to that expected for 
a dwarf-like photosphere.  {\it D05} estimated luminosities by performing
a bolometric correction to near-infrared $K$-band 
($\lambda \simeq 2.3 \mu$m) photometric data, after accounting for $K$-band 
veilings determined from their spectra; extinctions were determined by
comparing the $H-K$ colors to a typical value for a T Tauri star.
However, much of the flux detected from embedded young stars at visible and
near-infrared wavelengths has 
been scattered from their circumstellar environments.  The physical
nature of circumstellar dust grains (sizes, shape, composition),
distribution of material in disks and envelopes, and system
inclination all change how the photospheric flux is scattered
into our line of sight, changing both a star's brightness and color.
Comparisons of luminosities determined via different techniques differ 
by factors of 2-3, and we suggest 
this as a typical uncertainty.  In addition to this, stars with edge-on
disk orientations often have calculated luminosities that can be low by
factors of 10 to 100; the preferential short-wavelength scattering leads 
to artificially low extinction estimates.  With these uncertainties and 
possible systematic errors in mind, in Fig. 4 are shown all the 
Class I and flat-spectrum stars ($\alpha > -0.3$) in Tau-Aur, $\rho$ Oph, 
and Serpens observed in the {\it D05 } and the {\it WH04} surveys on 
a Hertzprung-Russell diagram.  The PMS evolutionary models of {\it 
Baraffe et al.} (1998) are shown for comparison.

Several points can be extracted from Fig. 4 regarding the masses of 
Class I and flat-spectrum stars.  First, the stars generally span a 
similar range of effective temperatures and stellar luminosities in all 
regions, though there is a slightly narrower range of temperatures in $\rho$ 
Oph.  Second, based on comparisons with the {\it Baraffe et al.} (1998) 
PMS evolutionary models, the combined distribution of stellar masses span 
from substellar to several solar masses.  This is similar to the 
distributions of Class II stellar masses in Tau-Aur and $\rho$ Oph 
({\it Kenyon and Hartmann}, 1995; {\it Luhman and Rieke}, 1999), while
little has been reported on the masses of Class II stars in the Serpens 
clouds.  Other studies corroborates these findings.
{\it Nisini et al.} (2005) determine masses spanning from 0.3 to 1.2
M$_\odot$ for 3 Class I stars in R CrA, based on spectral types determined
from moderate resolution spectra.  {\it Brown and Chandler} (1999)
determine masses of 0.2 - 0.7 M$_\odot$ for 2 Class I stars in Tau-Aur
based on disk kinematics under the assumption of Keplerian rotation.

\epsfig{file=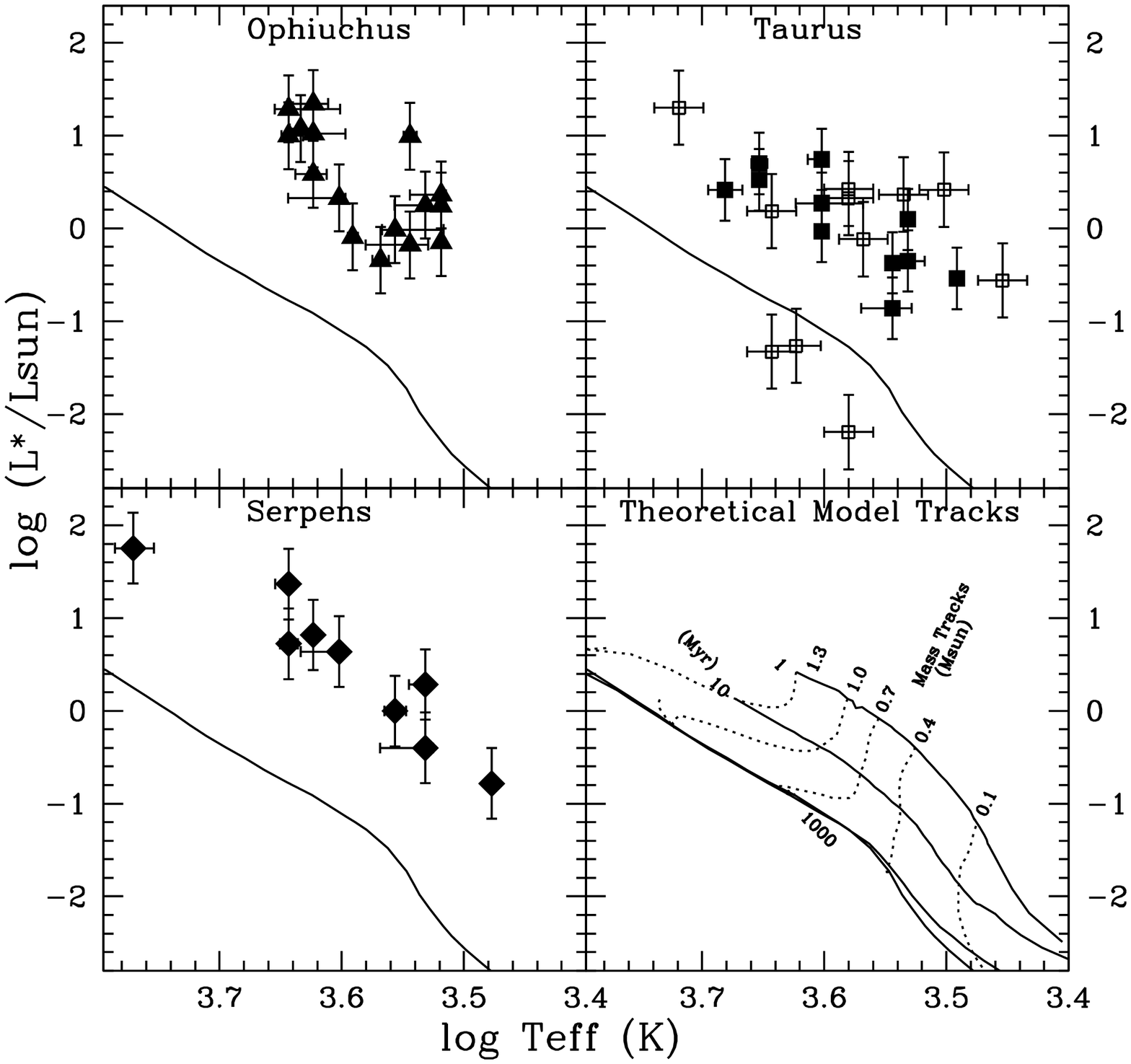, width = 3.1in} \noindent {\small Fig.
  4.  Stellar luminosities and effective temperatures of Class I and
  flat-SED stars ($\alpha > -0.3$) in $\rho$ Oph, Tau-Aur, and Serpens
  are shown on an H-R diagram.  Filled symbols are from D05 and unfilled
  symbols are from WH04.  The evolutionary models of {\it Baraffe et
  al.}  (1998) are also shown for comparison.} 
\bigskip

Apparently the majority of Class I stars have stellar masses that are similar
to those of T Tauri stars.  The present spectroscopic data does not support
the notion that the majority of the low luminosity ($L < 1 L_{\odot}$)
Class I stars are substellar.  This proposed resolution to the luminosity
problem can now be excluded.  Nevertheless, it is particularly notable that
there are several Class I stars which have masses that are close to or
below the substellar boundary (0.075 M$_\odot$; {\it Baraffe et al.},
1998).  {\it WH04} identified 3 Class I stars with spectral types of M5.5
or M6, and considered them candidate Class I brown dwarfs.  Two of these
stars, IRAS 04158+2805 and IRAS 04489+3042 were also observed and analyzed 
by {\it D05}.  The {\it D05} infrared spectra also indicate a M6 spectral 
type for IRAS 04158+2805, but yield a slightly earlier M4 spectral type 
for IRAS 04489+3042.  It is very encouraging that both optical and infrared 
spectra are yielding very similar results for these stars, and 
strengthens the case for the existence of a Class I object at or below
the substellar boundary.

On the other hand, some Class I stars which had been previously 
interpreted as accreting brown dwarfs from photometric data are now
revealed to be low mass stars instead.  For example, {\it Young et al.}
(2003) interpreted the very complete photometric data on the Class I star
IRAS 04385+2250 as evidence that it is a brown dwarf of only 0.01
$M_{\odot}$ ($\sim$ 10 $M_{\rm Jup}$) by assuming an accretion rate of
$2 \times 10^{-6} M_{\odot}$/yr.  As pointed out by {\it Kenyon
et al.} (1990; 1994), such an assumption implies that all low luminosity
($L < 1 L_{\odot}$) Class I stars are actually substellar.  However, 
{\it WH04} 
find that IRAS 04385+2250 (also known as Haro 6-33) has a spectral type of
M0, placing it squarely in the regime of low mass stars and not brown
dwarfs.  

\subsection{Stellar Ages}

In addition to providing stellar masses, comparisons of observationally
determined stellar properties with the predictions of evolutionary models
can provide useful age estimates.  Comparisons of Class II stars in nearby 
dark clouds consistently yield ages spanning from less than 1 to a few 
million years (e.g., see {\it Kenyon and Hartmann}, 1995; {\it Luhman and 
Rieke}, 1999).  If Class I stars are really the precursors to Class II 
stars, as their less evolved circumstellar environments suggest, then 
they should have younger ages.

As emphasized above, the calculated luminosities of Class I stars are
especially uncertain given the large extinctions, uncertain scattered 
light contributions, and continuum excesses; they are also occasionally 
subject to large systematic errors caused by orientation effects.  These
large uncertainties bear directly upon how well the stellar ages can be
determined, since low mass stars evolve primary along vertical evolutionary
tracks at ages less than a few $\times 10^{7}$ yr.  Nevertheless,
comparisons of the observed luminosities and temperatures with the
predictions of pre-main sequence evolutionary models, as shown in Fig. 4,
provide a large ensemble of ages estimates.
The range of ages is
broadest for Tau-Aur and narrowest for Serpens, though these regions have
the largest and smallest samples measured, respectively.  Several stars
in Tau-Aur appear to have unrealistically old ages (below the 
main-sequence), likely a consequence of the stellar luminosity being
severely underestimated because of an edge-on disk orientation (see 
{\it WH04}).  The absence of low luminosity stars in $\rho$ Oph and Serpens 
suggests they may be more difficult to identify in regions of high 
extinction. 

Despite these possible regional differences and large uncertainties,
calculating a median age in all 3 regions yields a consistent value
of $\sim 1$ Myr.  This is remarkably similar to the average age of
Class II stars in $\rho$ Oph and Tau-Aur; few are known in Serpens.
This suggests that most Class I stars are not systematically younger
than Class II stars.  Unfortunately the large uncertainties in the 
luminosity estimates of Class I stars, as well as current evolutionary
models at early ages ({\it Baraffe et al.}, 2002), limits the robustness of
this comparison at this time.  

Finally, we note that a comparison of the inferred stellar luminosities
with calculated bolometric luminosities for Class I stars suggests that 
in most cases studied here, the stellar luminosity is the dominant source 
of luminosity in the system ($L_{Star}/L_{Bol} > 0.5$).  Most Class I 
stars with detected photospheric features do not have accretion dominated
luminosities as had been initially 
proposed.  We caution, however, that the observational biases in the
sample studied here (revealed at $< 3 \mu$m, with moderate veiling or 
less) prevent extrapolation of this finding to Class I stars in general.
The most bolometrically luminous stars for which photospheric features
are detected are IRS 43 (L$_{bol}$ = 7.2 L$_\odot$) and YLW 16A (L$_{bol}$
= 8.9 $L_\odot$; see {\it D05}).  Thus it is not yet known if the most
luminous Class I stars (L$_{bol} > 10$ L$_\odot$) have accretion dominated
luminosities or are simply more massive stars.  


\subsection{Stellar Rotation}

Studies of stellar rotation at very young ages have revealed
clues regarding the evolution of angular momentum from the epoch
of star formation through to the young main sequence.
Conservation of angular momentum during the collapse
of a molecular core to form a low-mass star should lead to rotation
velocities near break-up 
($v_{break-up} = \sqrt($GM$_{star}$/R$_{star}) \sim 200$ km/s).
However, the small projected rotation
velocities of Class II stars ($v$sin$i \lesssim$ 20 km/sec; {\it
Bouvier et al.}, 1986, 1993; {\it Hartmann et al.}, 1986; {\it Stassun
et al.}, 1999; {\it Rhode et al.}, 2001; {\it Rebull et al.}, 2002;
see the chapter by {\it Herbst et al.})
show that angular momentum must be extracted quickly, on time scales
of $< 1-10$ Myr.  A number of theories have been proposed to
rotationally ``brake'' young stars.  One favored model for
involves magnetic linkage between the star and
slowly rotating disk material at a distance of several stellar radii
({\it K\"onigl}, 1991; {\it Collier Cameron and Campbell}, 1993; {\it Shu et
  al.}, 1994; {\it Armitage and Clarke}, 1996).  Initial observational
evidence 
supported this picture.  T Tauri stars without disks were found to
rotate somewhat more rapidly than stars with disks ({\em Edwards et
al.}, 1993; {\it Bouvier et al.}, 1993, 1995), which was interpreted
as evidence that disk presence keeps stars rotating at fixed
angular velocity while disk absence allows stars to conserve
angular momentum and spin up as they contract towards their
main sequence radii.  Since then, the observational case for disk locking
has become less clear-cut (e.g., {\it Stassun et al.}, 2001; but see {\it
  Rebull et al.}, 2004), while detailed theoretical and
magneto-hydrodynamical 
considerations suggest that disk locking in and of itself is unable to
extract sufficient amounts of angular momentum ({\it Safier}, 1998).
Strong stellar winds are one possible alternative (e.g., {\it Matt et al.},
2005).
\bigskip

\epsfig{file=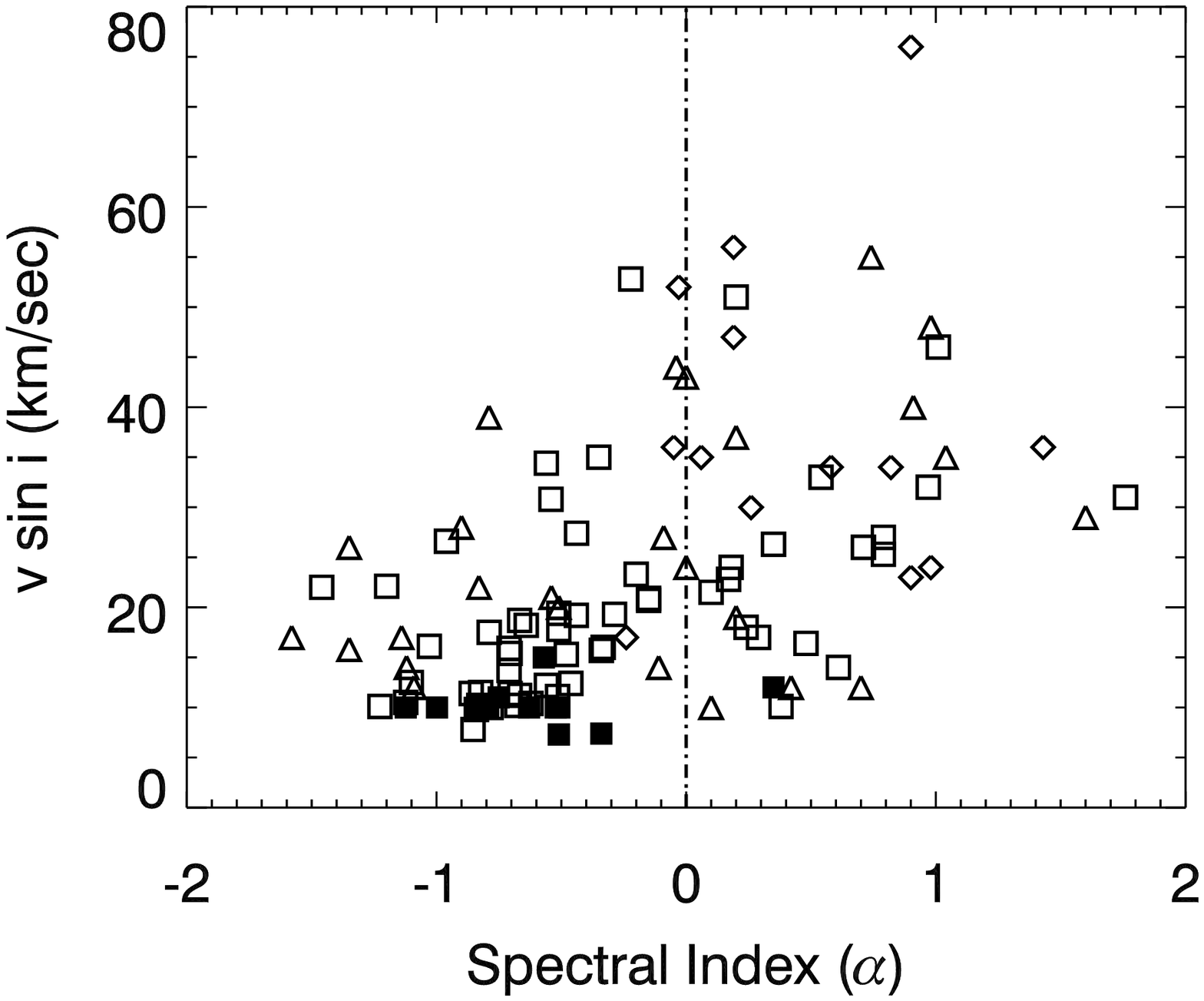, angle = 90, width = 3.0in}
\noindent
{\small Fig. 5.  Projected rotational velocity ($v$sin$i$) versus spectral
  index.  Triangles are stars in $\rho$ Oph, squares are stars in Tau-Aur
  and diamonds are stars in Serpens.  $v$sin$i$ measurements are shown as
  open symbols while upper limits are shown as filled symbols.  The dashed 
  vertical line separates Class I stars from Class II stars.  Data
  originally presented in {\it WH04}, {\it Covey et al.} (2005) and
  references therein.}
\bigskip

The uncertainties in our understanding of {\it how} angular momentum
is extracted from young stars provide motivation for determining
{\it when} it is extracted, since knowing the appropriate
time scale could help distinguish between proposed models.
The rotation velocities of
Class I stars as revealed by spectroscopic studies provide the earliest
measurements of stellar angular momentum; Fig. 5 shows the distribution of
$v$sin$i$ values for Class I and Class II stars in Tau-Aur, $\rho$ Oph, and
R CrA versus the evolutionary diagnostic $\alpha$.  The largest $v$sin$i$
value observed for a Class I star is 77 km/s, while the remainder have
$v$sin$i$ $\le 56$ km/s.  These values are only a few tenths of the typical
break-up velocity. 
Comparing Class I ($\alpha > 0.0$) to Class II ($\alpha < 0.0$) stars,
Class I stars have slightly higher rotation rates.  Although the
distributions of rotation rates are statistically different ({\it Covey et
  al.}, 2005), the difference in the mean is only a factor of
two.  The distributions are less distinct for any  particular region
(e.g., Tau-Aur, {\it WH04}), likely from smaller number statistics, though
global properties of a region could lead to correlated biases (e.g., age).
Although the distributions of Class II rotational velocities 
in some star forming regions have been shown to be statistically different
(e.g., Orion versus Tau-Aur; {\it Clarke and Bouvier}, 2000; {\it White and
Basri}, 2003), the evidence for this at the Class I stage is still
tentative ($\sim 2\sigma$); {\it Covey et al.} (2005) found Tau-Aur to
have the lowest mean observed rotation velocity for the three regions in
their study (30.1 km/s versus 31.1 km/s in $\rho$ Oph and 36.8 km/s in
Serpens).  These comparisons are likewise limited by small number
statistics.  Overall, the observational evidence demonstrates that Class I
stars are rotating somewhat more rapidly than Class II stars, but at rates
that are well below break-up velocities.  If Class I stars are indeed
in the main phase of mass accretion (Section 4.3), this implies that
angular momentum is removed concurrently with this process.

\subsection{Circumstellar Accretion}

If Class I stars are to acquire the majority of their mass (e.g., 0.6
M$_\odot$) on a timescale of $\sim 2 \times 10^5$ yr, they must have
time-averaged mass accretion rates that are $\sim 3 \times 10^{-6}$
M$_\odot$/yr, assuming a simple spherical infall model (e.g., {\it
  Hartmann}, 1998).  For comparison, this mass accretion rate is at least 2
orders of magnitude larger than what is typically observed for T Tauri
stars (e.g., {\em Gullbring et al.}, 1998).  The newly available high
dispersion spectra of Class I stars permit measurements of the mass
accretion rate (from the disk onto the star), by 2 independent methods.
The first of these comes from measurements of optical excess emission in
high dispersion optical spectra under the assumption that the liberated
energy is gravitational potential energy (see the chapter by {\it Bouvier
  et al.}).  Unfortunately there remain considerable uncertainties in
measuring the total liberated energy, which typically requires a large
bolometric correction from an optical measurement; the majority of the
accretion luminosity is emitted at ultra-violet wavelengths.  Additionally,
estimating the potential energy requires estimates of stellar and inner
disk properties, which have large uncertainties themselves.  Nevertheless,
by calculating mass accretion rates for Class I stars following the same
assumptions used for T Tauri stars, many of these systematic uncertainties
can be removed, thereby permitting a more robust comparison if the same
accretion mechanism applies.

{\it WH04} have measured optical excess emission at 6500 \AA\, for 11 Class
I stars, and several borderline Class I/II stars, all within the Tau-Aur
star forming region.  These measurements, along with a sample of excess
measurements of accreting T Tauri stars from {\it Hartigan et al.} (1995;
as compiled in {\it WH04}), are shown in Fig. 6.  Similar to the T Tauri
stars, the Class I stars have continuum excesses that range from not
detected ($<0.1$) to several times the photosphere; in the general case, 
their optical emission is not dominated by accretion luminosity.
Class I stars have veiling values that are only modestly larger in the
mean ($\times 1.3$) than those of Class II stars.  {\it WH04} proceed to
convert these continuum excesses to mass accretion rates, and find 
values of a few $\times 10^{-8}$ M$_\odot$/yr, which are again similar to
those of Class II stars.  Further, by accounting for other components to
the star's luminosity, they find that the accretion luminosity only
accounts for $\sim 25$\% of the bolometric luminosity, on average.

A second measure of the mass accretion rate comes from emission line
luminosities.  Emission-line studies, in combination with radiative
transfer models of circumstellar accretion, suggest that many of the
permitted lines originate in the infalling magnetospheric flow ({\it
  Hartmann et al.}, 1994; {\it Muzerolle et al.}, 1998), and that the line
strengths are proportional to the amount of infalling mass.  {\it Muzerolle
  et al.} (1998a, 1998b) demonstrated this to be true for the Ca II
infrared triplet and Br$\gamma$ by correlating these emission-line
luminosities with mass accretion rates determined from blue excess
emission.  As emphasized by the authors, accurate corrections for
extinction and scattered light are critical for this.  The near-infrared
emission-line Br$\gamma$ is of special interest in the study for Class I
stars since the high extinction often inhibits observations at shorter
wavelengths.  Using the Br$\gamma$ correlation, {\it Muzerolle et al.}
(1998) found that the Br$\gamma$ luminosities of Class I stars, with
assumed stellar properties, are similar to those of Class II stars.  The
implication is that they have similar mass accretion rates.
\bigskip

\epsfig{file=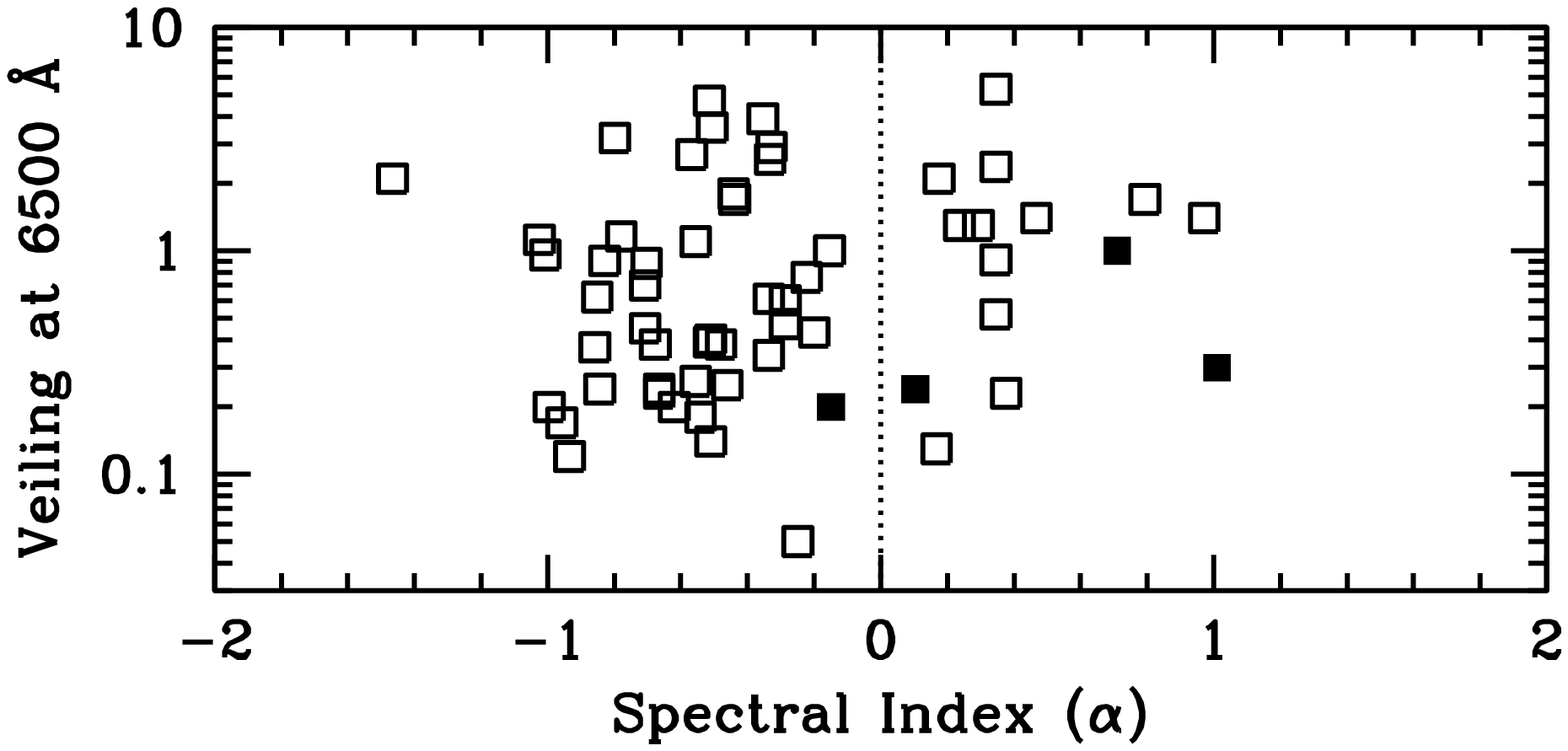, width = 3.1in}
\noindent
{\small Fig. 6.  Optical veiling versus spectral index for stars in Tau-Aur.
  Measurements are 
  from {\it WH04} and {\em Hartigan et al.}
  (1995).  Veiling measurements are shown as open symbols while upper
  limits are shown as filled symbols.  The dashed vertical line separates
  Class I stars from Class II stars.}
\bigskip

Fig. 7 shows a compilation of logarithmic Br$\gamma$ luminosities from {\it
Muzerolle et al.} (1998) and D05 (also includes measurements from {\it
  Luhman and Rieke}, 1999; {\it Folha and Emerson}, 1999; {\it Doppmann et
  al.}, 2003), 
plotted versus spectral index.  As with optical excess emission, the
Br$\gamma$ luminosities of Class I stars span a similar, though slightly
broader range than the Class II stars, but are larger in the average
by a factor of a few; the distributions are different at approximately the
3$\sigma$ level according to a K-S test.  Much of the difference between
the Class I and Class II stars appears to be driven by stars in the
$\rho$ Oph region, where the Br$\gamma$ luminosities of Class I stars are
systematically larger than those of Class II stars by a factor of $\sim
5$ in the mean; the distributions are different at the $\sim 2\sigma$
  level, or $\gtrsim 3\sigma$ if the low Br$\gamma$ luminosity (-4.58),
  $\alpha = 0.0$ star GY21 is considered a Class II.  Stars in Tau-Aur show
no difference between the 2 classes ($< 1\sigma$).

Conversion of these luminosities to mass accretion rates leads to values
for Class I stars that are similar to Class II stars, and corroborates the
initial study of {\it Muzerolle et al.} (1998).  The largest mass accretion
rates 
are $\sim 10^{-7}$ M$_\odot$/yr, and many of these are in the $\rho$ Oph
star forming region.  The larger mean accretion luminosities in $\rho$ Oph is
consistent with its larger mean near-infrared excess for Class I stars
relative to Class II stars ($<r_{K}> = 2.2$ versus $0.94$), compared with
other regions.  Overall it appears that the mass accretion rate during
the majority of the Class I phase is similar to that of T Tauri stars, and
$1-2$ orders of magnitude less than the 
envelope infall rates inferred from SED modeling (few$\times 10^{-6}$
M$_\odot$/yr).  We note that there is tentative evidence that 
the mass accretion rate is extremely time variable during the embedded 
phase.  As one example, the borderline Class I/II star IRAS
04303+2240 changed its mass accretion rate dramatically ($>4\times$)
during 2 observational epochs (Fig. 2).  Little observational work has been
done to characterize the amplitudes or timescales of candidate protostar 
variability.

\epsfig{file=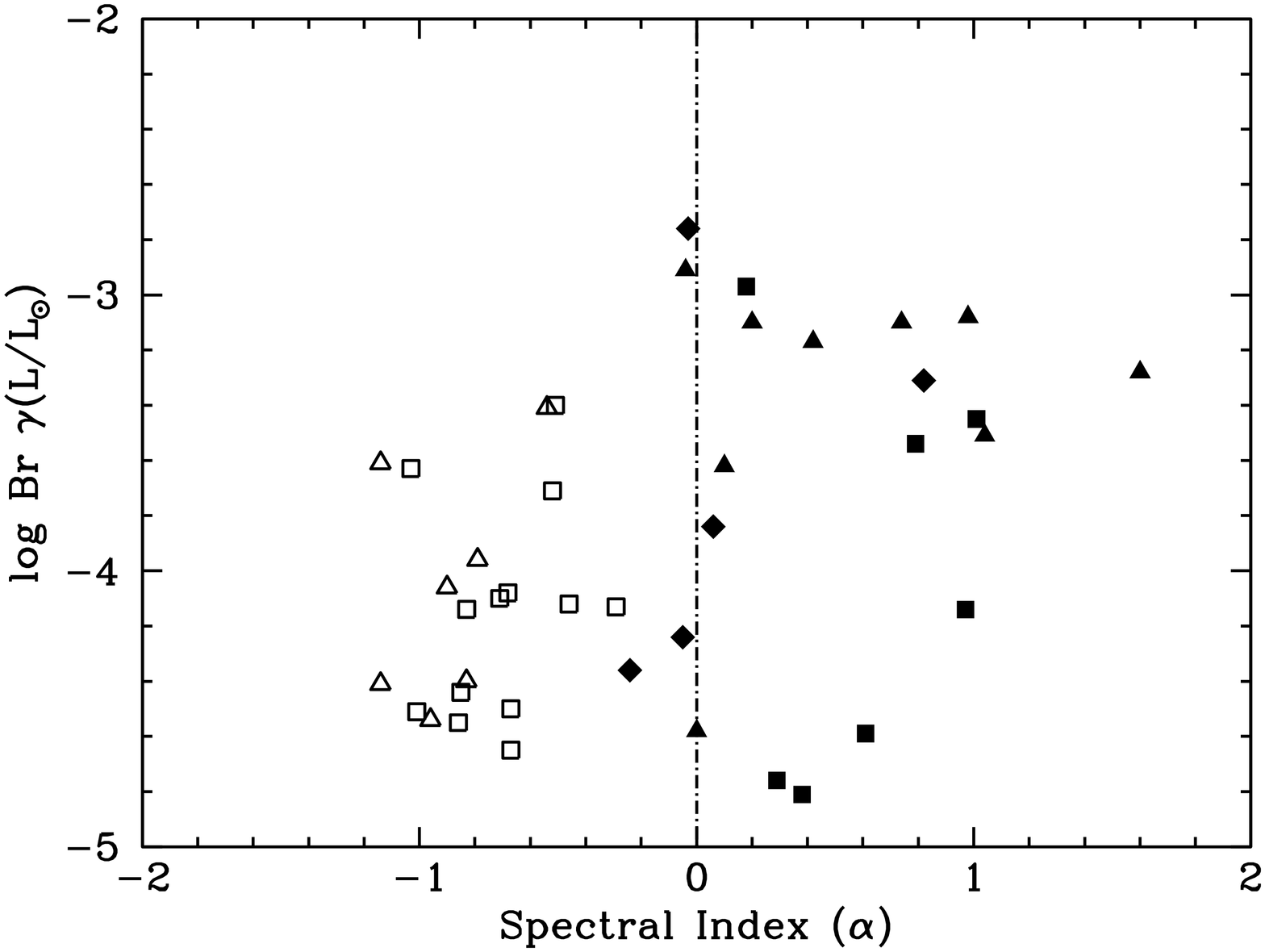, angle = -90, width = 3.2in}
\noindent
{\small Fig. 7.  Br$\gamma$ luminosity versus spectral index.  Triangles 
 are stars in $\rho$ Oph, squares are stars in Tau-Aur and Diamonds are
 stars in Serpens.  Filled symbols are from {\it D05} while open symbols
 are from {\it Muzerolle et al.} (1998).  The dashed vertical line
 separated Class I stars from Class II stars.}

\subsection{Jet Emission}

Optically thin forbidden emission-lines are believed to originate in
an outflowing jet or wind.  Their intensity is expected to be
directly proportional to the amount of material being funneled along the
jet, as viewed through the slit of the spectrograph.  The luminosity of
these emission lines can therefore be used to estimate the mass outflow
rate in a young stellar jet (see the chapter by
{\it Bally et al.}).  Since jets are believed to be powered by
circumstellar accretion, the mass outflow rate should correlate with
the mass accretion rate.

In Fig. 8 are shown equivalent width measurements of the forbidden line
[SII] 6731 \AA\, for stars in Tau-Aur versus spectral index.
Measurements are from WH04 and {\it Hartigan et al.} (1995; as compiled in 
{\it WH04}).  Unlike the optical excess and Br$\gamma$ luminosity accretion
diagnostics, which are only slightly enhanced among Class I stars 
relative to Class II stars, Class I stars systematically have larger
[SII] equivalent widths by roughly a factor of 20 in the mean.  The
implication is that Class I stars power much more energetic outflows than
Class II stars.  {\it Kenyon et al.} (1998) found similar results based on
some of the same stars presented here.

However, as emphasized by {\it WH04}, many of the Class I stars they observed  
show signatures of having an edge-on disk orientation.  In such a case, 
the emission-line region may be more directly observable than the partially 
embedded central star is.  The preferentially attenuated continuum flux will 
consequently produce artificially large equivalent width values, and biased 
mass outflow rates.  Without accurate geometric information for these stars,
however, it is difficult to tell the significance of this bias in the average
case.  If these larger forbidden line equivalent widths indeed correspond to 
larger mass outflow rates, perhaps it is because their spatially extended 
location make them a better tracer of the time-averaged mass outflow rate.
As is considered below, the mass accretion rate (and corresponding mass
outflow rate) for Class I stars could be T Tauri-like for the majority 
of the time, with occasional large outbursts.  In such a case, Class I
stars would then have a larger time-averaged mass accretion and mass
outflow rates.  The alternative to this, in the absence of any significant
continuum attenuation bias, is that the ratio of mass loss to mass
accretion rate is 
dramatically different between Class I and Class II stars ($>10\times$),
possibly suggesting a different accretion mechanism.
\bigskip

\epsfig{file=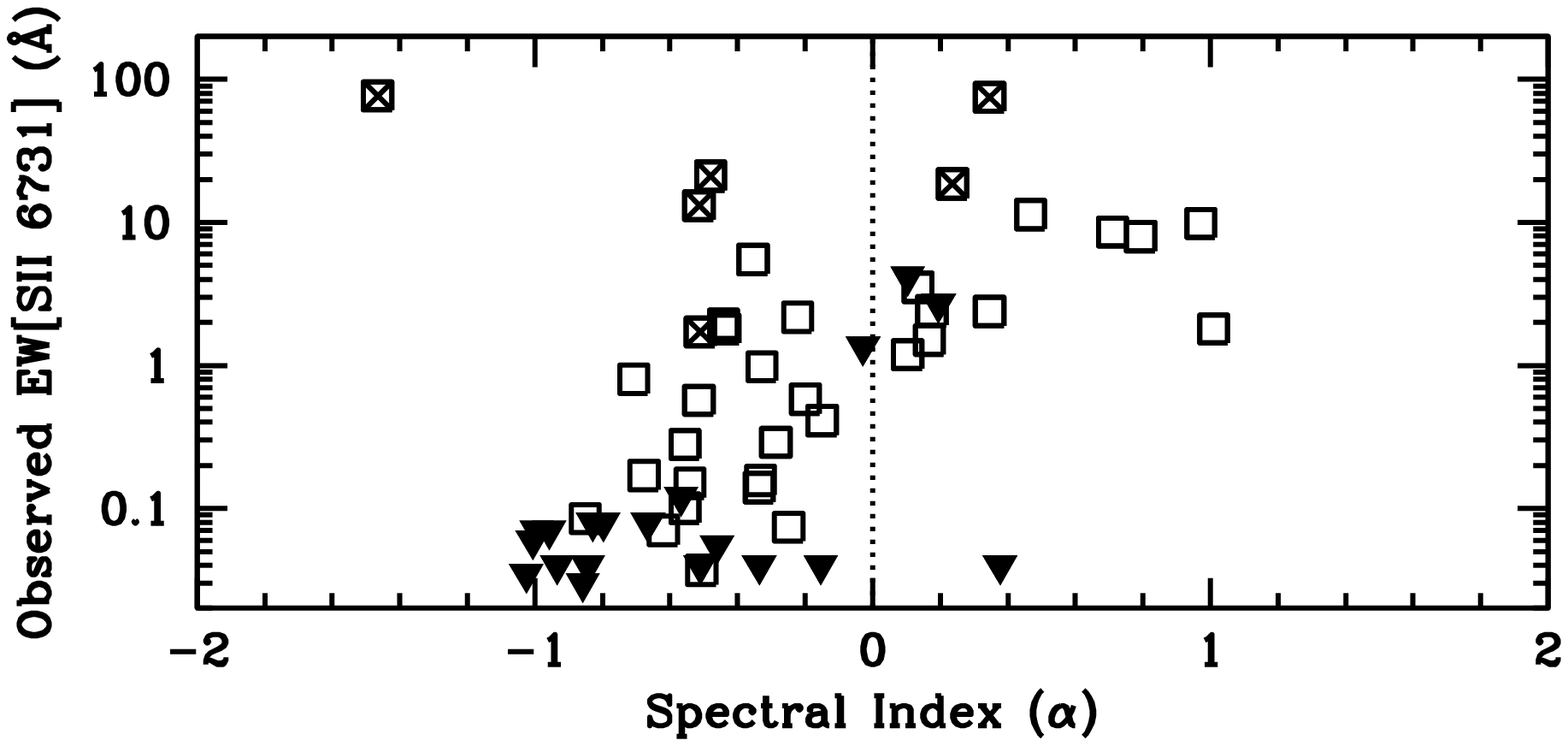, width = 3.1in}
\noindent
{\small Fig. 8.  Equivalent width measurements of [SII] 6731 \AA\, versus
  spectral index for stars in Tau-Aur.  Measurements are from {\it WH04} 
  and {\it Hartigan et al.} (1995).  Detections
  are shown as open squares while upper limits are shown as
  filled triangles; stars with an x have a known or suspected edge-on
  orientation.  The dashed vertical line separates Class I stars from Class
  II stars.}
\bigskip

{\centering
\section{\textbf{IMPLICATIONS FOR CLASS CLASSIFICATION SCHEMES AND
    FORMATION THEORY}}
}

A broad comparison of the properties derived for Class I stars and
Class II stars from spectroscopy reveals some surprising similarities 
and differences.  Class I stars appear to occupy the same ranges of 
effective temperature and (photospheric) luminosity as Class II
stars; applying our current theoretical understanding of PMS
evolution to these observations implies that Class I stars have
similar stellar masses as Class II stars.  If Class I stars are 
indeed precursors to T Tauri stars, this means that by the Class 
I stage a young star has accreted the majority of its final stellar 
mass.  Although the inferred stellar luminosities of Class I stars 
are, on average, similar to those of Class II stars, implying similar 
ages, the large uncertainties and systematic biases in these 
estimates prevent strict age comparisons at this time.  
Class I stars appear to be more
rapidly rotating, on average, than Class II stars, though there are
many Class I stars with low projected rotational velocities.
Spectroscopic indicators of mass accretion, such as optical veiling
and Br $\gamma$ luminosity, do appear slightly elevated in Class I
stars relative to Class II stars, but still well below predicted
 mass infall rates ($\sim 10^{-6}$
M$_\odot$/yr).  Although Class I stars have larger forbidden line
emission strengths, implying larger mass outflow rates, there is a
yet unaccounted for continuum attenuation bias in these measurements.  Here
we use the combined results to investigate possible regional differences
among Class I stars, to assess whether Class I stars are being properly
classified, and to improve our understanding of how mass is acquired 
during the main phase of mass accretion.

\medskip
\subsection{Are There Regional Differences Among Class I Stars?}

Initial studies of Class I stars suggested that their properties may
differ in different regions.  {\it Kenyon et al.} (1990) noted that
although the Tau-Aur and $\rho$ Oph clouds both contain similar
numbers of Class I stars, $\rho$ Oph contains many more with $L_{\rm
bol} > 10 L_{\odot}$.  If stars in both regions have similar stellar
masses, then this luminosity difference should translate into $\rho$
Oph stars having mass accretion rates 3 -- 10 times higher than those 
in Tau-Aur (and correspondingly larger mass infall rates, if the
accretion is steady-state).
This is expected according to classical star formation theory ({\it Shu}, 
1977), which predicts that the infall rate should scale as the cube of 
the isothermal sound speed.  The warmer gas in $\rho$ Oph, relative to 
Tau-Aur ({\it Myers and Benson}, 1983), should consequently yield large 
mass infall rates, and larger time-averaged mass accretion rates.  
However, more recent work suggests that cloud turbulence may primarily 
set the initial infall rates (e.g., {\it Mac Low and Klessen}, 2004), 
and possibly even the initial mass function (e.g., {\it Goodwin et al.},
2004), and the 
resulting binary fraction and rotational distribution ({\it Jappsen 
and Klessen}, 2004).  Searching for possible differences in the stellar 
and accretion properties of stars produced in regions with different
global properties (temperature, turbulence, density) can therefore help 
distinguish between proposed scenarios for mass assembly and early
evolution.

The distributions of effective temperatures shown in Fig. 4 indicate
that there are no significant differences in the masses of Class I
stars in either the Tau-Aur, $\rho$ Oph, or Serpens star forming
regions ($< 1\sigma$, according to K-S tests). 
While the stellar luminosities and ages are also similar among the 3
regions (given their large uncertainties), Serpens is somewhat
distinct in that all of its Class I and flat-SED members appear coeval
at an age younger than 1 Myr, as would be expected for 
bona-fide protostars.  The larger scatter in stellar luminosity in
$\rho$ Oph and Tau-Aur is not well understood, though some apparently
low luminosity stars are a consequence of their edge-on disk
orientation (Section 3.2).  

The analysis of accretion diagnostics in Section 3.4 partially supports a
scenario in which Class I stars in $\rho$ Oph are accreting at higher
rates than those in Tau-Aur.  The Br$\gamma$ luminosities of 
Class I stars in $\rho$ Oph are systematically larger than those in Tau-Aur,
implying larger mass accretion rates.  This is also supported, though less
directly, with the larger near-IR veiling and higher bolometric luminosities
of Class I stars in $\rho$ Oph relative to Tau-Aur ({\it D05}).
In a case study of the luminous ($L_{bol} =$ 10 L$_\odot$) protostar
YLW 15, {\it Greene and Lada}, (2002) determine that 70\% of the star's
luminosity is due to mass accretion and infer a rate of $2 \times 10^{-6}
M_{\odot}$/yr.  At least in this one case, the disk accretion rate
appears consistent with the mass infall rate inferred from envelope models
(though these infall rates are derived primarily from Class I stars in 
Tau-Aur, because of less confusion with cloud material in that region).  
Given the small number and broad range of Br$\gamma$ luminosities for 
Class I stars in Serpens, this distribution is consistent with the 
distributions of either $\rho$ Oph or Tau-Aur Class I stars.

Finally, the present data do not reveal any notable differences in the 
distributions of rotation velocities, or angular momenta,
of embedded protostars in different regions.  {\it 
Covey et al.} (2005) found that Class I and flat spectrum stars in 
Serpens had a somewhat larger mean $v$sin$i$ rotation velocity than 
those in Tau-Aur or $\rho$ Oph, but this difference is not statistically 
significant ($< 2\sigma$).  We caution that any orientation bias
present in the samples studied (e.g., edge-on disk systems), will also 
bias the distribution of projected rotational velocities.

\subsection{Are Class I Stars Properly Classified?}

In the traditional classification scheme, Class I stars are true protostars
-- stellar embryos surrounded by an infalling envelope -- while Class II 
stars are pre-main sequence stars surrounded by circumstellar disks only.
Here we consider the ability of popular evolutionary diagnostics to
unambiguously  distinguish between these 2 classes.
Radiative transfer models of still-forming stars find that the SED shape 
typically used to distinguish Class I and Class II stars (as parametrized by 
T$_{bol}$ and $\alpha$) has an important dependence on the orientation of
the disk and envelope relative to the observer's line of sight ({\it Kenyon  
et al.}, 1993a, 1993b; {\it Yorke et al.}, 1993; {\it  Sonnhalter et al.},
1995;  
{\it Whitney et al.}, 2003, 2004).  As an example, the models of {\it
Whitney et al.} (2003) show that mid-latitude ($i \sim 40^{\circ}$) Class
I stars have optical, near- and mid-infrared characteristics similar to
those of more edge-on disk ($i \sim 75^{\circ}$) Class II stars.  The
effects of edge-on disk orientation are most severe for evolutionary
diagnostics determined in the near- and mid-infrared such as the $2-25\,
\mu$m spectral index.  Bolometric temperatures are also biased, but less 
so, while diagnostics based at much longer wavelengths, 
such as the ratio of sub-millimeter to bolometric luminosity ({\it Andr\'{e} 
et al.}, 1993), are the least affected.  Unfortunately longer wavelength 
SEDs are not yet available for Class I stars in many star forming region.
Consequently, we conclude that current samples of Class I stars defined by 
either spectral index or bolometric temperature are contaminated with at
least a few edge-on disk Class II stars.

Other observable characteristics, however, can be helpful in identifying
Class II stars that have been mistakenly classified as Class I stars due to 
orientation effects.  The radiative transfer models of {\it Whitney et al.} 
(2003) show that edge-on Class II stars are nearly 5 times fainter than `true' 
Class 
I stars with the same value of $\alpha$, suggesting that misclassified edge-on 
systems should appear significantly lower in an H-R diagram.  {\it WH04} 
identified several likely disk edge-on systems in their sample of optically 
revealed Class I stars in Tau-Aur, several of which (but not all) appear 
under-luminous relative to other cluster members.  Unfortunately the large
luminosity spread of Class I stars inhibit identifying edge-on disk systems
based on this criterion alone, unless the system is almost precisely
edge-on (e.g. HH 30).  Column-density sensitive spectral features
(e.g., Si at $9.7 \mu$m; {\it Kessler-Silacci et al.}, 2005) or high spatial 
resolution imaging may provide less ambiguous orientation information.

The presence of spatially extended envelope material, as determined from 
image morphology at infrared and millimeter wavelengths, has been proposed as
a more direct way to constrain the evolutionary Class.  Such features are 
only expected during the main accretion phase.  Based on
criteria put forth by {\it Motte and Andre} (2001), only 58\% (15/26) of the
Class I stars in Tau-Aur are true protostars.  The remaining 42\% (11 stars)
have envelope masses $\lesssim 0.1$ M$_\odot$ and are spatially unresolved
at 1.3 mm wavelengths (referred to as ``unresolved Class I sources'' in
{\it Motte and Andre}, 2001). {\it Motte and Andre} (2001) suggest that
these stars are 
more likely transitional Class I/II stars or highly reddened Class II
stars (e.g., edge-on disk systems).  The complementary near-infrared morphology 
survey by {\it Park and
Kenyon} (2002) supports the claim that these stars are not bona fide Class I
stars.  However, we note that the morphological criteria used in these
studies do not account for the luminosity and mass of the central star.
For example, IRAS 04158+2805 may appear more evolved and point-like
because it is a lower luminosity Class I brown dwarf with a smaller disk
and envelope.

{\it Andr\'e and Montmerle} (1994) present a similar morphological study
based 1.3 mm continuum observations of Class I and Class II stars in the 
$\rho$ Oph star forming region.  They found 
that Class I and Class II stars, as classified by the $2.2 - 10\, \mu$m 
spectral index, have similar 1.3 mm flux densities.  Class I stars, however, 
were more often spatially extended, consistent with a significant envelope 
component, though of relatively low mass ($\lesssim 0.1$ M$_\odot$).  Thus, 
it appears that a much smaller fraction of Class I stars in $\rho$ Oph,
relative to Tau-Aur, are candidate misclassified Class II stars.
Nevertheless, their low envelope masses imply that they have already
acquired the majority of their stellar mass (discussed below), like Class
II stars.  Complementary comparisons of Class I and Class II stars in the
Serpens and R CrA star forming region have not yet been carried out.

Based on this mostly indirect evidence, we conclude that between one-third
and one-half of the Class I stars in Tau-Aur are candidate misclassified
Class II stars; the emission-line profiles and image morphology suggests
that in some cases the misclassification is caused by a nearly edge-on
orientation.  There is less evidence for significant misclassification in
other regions (e.g., $\rho$ Oph).

\subsection{Are Class I Stars in the Main Accretion Phase?}
 
Although the absolute values of the circumstellar disk accretion rates 
have large systematic uncertainties, the rates inferred for
Class I stars and Class II stars, under the same assumptions, are
similar.  However, these values are typically 1-2 orders of magnitude 
less than both the envelope infall rates inferred from SED modeling of
Class I stars
(e.g., few $\times 10^{-8}$ M$_\odot$/yr vs. few $\times 
10^{-6}$ M$_\odot$/yr) and the time-averaged accretion rate necessary 
to assemble a solar mass star in a few $\times 10^5$ years.  Here we 
explore possible ways to reconcile this apparent discrepancy.

The first possibility to consider is that either the disk accretion 
rates or the mass infall rates are wrong, or both.  Given the large
uncertainty in determining the total accretion luminosity from an
observed excess, which is roughly an order of magnitude (see e.g., 
{\it Gullbring et al.}, 1998; {\it WH04}), the average disk accretion rate
could be as large as $10^{-7}$ M$_\odot$/yr.  Much larger disk accretion
rates would invoke statistical problems since classical T Tauri stars 
are accreting at this rate as well, for 1-10 Myr, and would 
consequently produce a much more massive population than what is 
observed.  Larger rates would also be inconsistent with emission-line
profile analyses ({\it Hartmann et al.}, 1994; {\it Muzerolle et al.},
1998).  Assessing possible errors in the mass infall rates is more 
challenging since most are not determined directly from kinematic 
infall signatures.  Instead, they are primarily set by the density 
of the envelope material; denser envelopes yield higher
mass infall rates and redder SEDs.  However, effects such as 
orientation ({\it Whitney et al.}, 2003) and disk emission ({\it
Kenyon et al.}, 1993a; {\it Wolf et al.}, 2003) can also shift the
SED towards redder wavelengths, if unaccounted for.  Using
sophisticated envelope plus disk models combined with spatially 
resolved images to constrain orientation, {\it Eisner et al.} 
(2005) and {\it Terebey et al.} (2006) nevertheless find that mass
infall rates of a few $\times 10^{-6}$ M$_\odot$/yr still provide
the best fits to the SED and image morphology.  How low these infall 
rates could be and still provide reasonable fits is unclear; a 
factor of $\sim 10$ decrease in the assumed infall rate could potentially 
reconcile the discrepancy, if disk accretion rates are correspondingly
increased by a factor of 10.  It is important to keep in mind, as
highlighted by {\it Terebey et al.} (2006), that the amount of envelope
material which actually reaches the star may be only one-forth of the
infalling mass because of mass lost to stellar jets/winds and companions.
With all this in mind, we conclude that it is possible to reconcile the 
infall/disk accretion rate discrepancy based on systematic errors and model
assuptions alone.  However, since the current best estimates strongly favor
values that are $\sim 2$ orders of magnitude descrepant, we will also
consider other possibilities for reconciling these rates.

One possibility, as first suggested by {\it Kenyon et al.} (1990), is that
the infalling envelope material is not transferred to the star via disk
accretion in a steady-state fashion.  Instead, the accreting envelope mass
accumulates in the circumstellar disk until it becomes gravitationally
unstable (e.g., {\it Larson}, 1984) and then briefly accretes at a prodigious
rate ($\sim 10^{-5}$ M$_\odot$/yr; see {\it Calvet et al.}, 2000).  
This scenario is consistent with the
small population of young, often embedded stars which dramatically
increase their luminosity for a few years to a few centuries (e.g., FU Ori,
{\it Hartmann and Kenyon}, 1987; V1647 Ori, {\it Brice\~no et al.}, 2004).
If Class I stars intermittently accrete at this rate, they must spend
5-10\% of their lifetime in the high accretion state to achieve typical 
T Tauri masses within 1 Myr.  Statistically, 5-10\% of Class I stars should 
then be
accreting at this rate.  The sample of Class I stars with mass accretion
rates is now becoming large enough to suggest a possible problem with these
expected percentages; none appear to accrete at this high of a rate (Section 3.4). 
However, there is a strong observational bias in that stars accreting at
this rate are likely to be too heavily veiled, at both optical and infrared
wavelengths, to identify photospheric features from which the amount of
excess can be measured.  Indeed, several stars observed by {\it WH04} and 
{\it D05} are too veiled to measure mass accretion rates.  L1551 IRS 5, 
for example, which is the most luminous Class I star in Tau-Aur, has been 
proposed to be a young star experiencing an FU Ori-like outburst ({\it
  Hartmann and Kenyon}, 1996; {\it Osorio et al.}, 2003).  Without a more 
accurate measure of its stellar properties this is difficult to confirm;
its larger luminosity could be a consequence of it being a somewhat more
massive star.

An independent test of the episodic accretion hypothesis is the relative
masses of Class I disks compared to Class II disks.  If the envelope
material of Class I stars is accumulating in their circumstellar disks,
they should be more massive than Class II stars.  {\it WH04}
investigated this using 1.3 mm continuum observations from
{\it Beckwith et al.} (1990), {\it Osterloh and Beckwith} (1995), and {\it
  Motte and Andr\'e} (2001), and restricted to beam sizes of $11-12$ $''$
to avoid contamination from envelope emission of Class I stars.  This 
comparison showed that the 1.3 mm flux densities of Class I and Class II
stars in Tau-Aur are indistinguishable, implying similar disk masses if the
Class I and Class II disks have similar dust opacity and dust temperature
({\it Henning et al.}, 1995).  However, {\it Andrews and Williams} (2005)
drew a different conclusion based on submillimeter observations at 450
$\mu$m and 850 $\mu$m (with beam sizes of 9'' and 15'', respectively).
They showed that the distribution of sub-millimeter flux densities and disk
masses of Class I stars are statistically different from those of Class II
stars (being more massive), though Class I and Class II samples
nevertheless span the same range of disk masses.  Unfortunately biases
introduced by stellar mass, multiplicity, envelope emission, and low
spatial resolution evolutionary diagnostics, inhibit robust comparisons of
these samples.  We conclude there is at most marginal evidence for Class I
stars having more massive disks than Class II stars, as would be expected
if they undergo FU Ori-like outbursts more often than Class II stars.

Given the overall similarities of Class I and Class II stars, {\it WH04} put
forth the still controversial suggestion that many (but not all) Class I
stars are no longer in the main accretion phase and are much older than
traditionally assumed; {\it WH04} focus their study on Class I stars in
Tau-Aur, where the case for this is most compelling.  This proposal does
not eliminate the luminosity problem for bona-fide Class I stars, but
minimizes the statistical significance of it in general.  Support for
this idea originates in the known biases introduced by current
classification criteria which are inadequate to unambiguously identify
young stars with infalling envelopes.  The two largest biases are the low
spatial resolution mid-infrared measurements upon which most SEDs are based
and the effects of an unknown orientation on the SED.  These biases likely
explain why $\sim 42$\% of stars classified as Class I stars in Tau-Aur do
not appear to be bona fide protostars (Section 4.2).  Indeed, some authors have
claimed that Class I stars like IRAS 04016+2610 and IRAS 04302+2247 have 
morphologies and kinematics that are better described by a rotating
disk-like structure ({\it Hogerheijde and Sandell}, 2000; {\it Boogert et
al.}, 2002; {\it Wolf}, 2003) than a collapsing envelope model ({\it Kenyon
  et al.}, 1993b; {\it Whitney et al.}, 1997), though more recent work
still favors massive envelopes (e.g., {\it Eisner et al.}, 2005).  However,
the limitation of all of these models is that 
they only account for the spatial distribution of circumstellar material,
which can be confused with diffuse cloud emission ({\it Motte and Andr\'e},
2001) or companion stars with $\sim 10^3$ AU separations ({\it Haisch et
  al.}, 2004; {\it Duch\^{e}ne et al.}, 2004).  A convincing case for a
massive infalling envelope can only be established by spatially mapping 
molecular line profiles and accounting for the effects of outflows and
rotations ({\it Evans}, 1999).  Currently the Class 0 star IRAS 04368+2557 
(L1527) is the only star in Tau-Aur that has been shown to retain a 
massive extended envelope with unambiguous evidence for infall ({\em 
Gregersen et al.}, 1997).

In regions outside Tau-Aur, there is less evidence as well as less motivation
for Class I stars being older than presumed and past the main phase of mass
accretion.  As discussed in Section 3.4, the higher disk accretion rates of
many Class I stars in $\rho$ Oph, for example, are within a factor of $\sim
10$ of predicted mass infall rates, and thus easier to reconcile given 
current uncertainties in observations and models assumptions.  Additionally,
there is less evidence that these Class I stars are misclassified Class II
stars, compared with Tau-Aur Class I stars.  However, we strongly caution
that it is not yet possible to tell if the apparent differences between the
Class I population in Tau-Aur and other regions reflects real differences 
in their evolutionary state or is simply a consequence of Tau-Aur being a
lower density environment and its members being more optically revealed.
One important similarity of Class I stars in all star forming regions is
their relatively low mass envelopes (e.g. {\it Andr\'e and Montmerle},
1994; {\it Motte and Andr\'e}, 2001), suggesting that at this phase they
have already acquired the majority of their stellar mass.

If the ages of Class I stars are indeed as old as T Tauri stars ($\gtrsim
1$ Myr) as the comparisons tentatively suggest (Section 3.2), there is a
potential dynamical timescale problem.  In such a case the envelope is
surviving nearly a factor of 10 longer than its dynamical collapse
timescale, which seems unlikely.  However, it is well known that there is
nearly an order of magnitude spread in the {\em disk} dispersal timescale
of Class II stars (e.g., {\it Hillenbrand et al.}, 1998); a similar spread
in the envelope dispersal timescale seems plausible.  One possibility for
generating a large spread in the envelope dispersal timescale is that in
some cases the envelopes are replenished.  Recent simulations of cluster
formation (e.g., {\it Bate et al.}, 2003) suggest that even after
the initial phase of mass accretion, a young star continues to 
dynamically interact with the cloud from which it formed, and in some
cases even significantly increase its mass.  Thus some embedded stars
could in fact come from an older population.  These would be difficult to 
distinguish from younger stars in their initial main accretion phase based
on circumstellar properties alone.  More accurately determined age
estimates is likely the best way to test this intriguing hypothesis.

Summarizing, we find that in most cases the disk accretion rates of Class
I stars are well below predicted envelope infall rates.  In some
cases this may be a consequence of misclassification.  In the more general
case, it implies that if the envelope material is indeed infalling, it is
not transferred to the star efficiently (e.g., {\it Terebey et al.}, 2006)
or at a steady rate (e.g., {\it Kenyon et al.}, 1990), or both.  While it is
  known that some 
young stars dramatically increase in brightness, presumably due to enhanced
accretion (e.g., FU Ori), the idea that this is process by which stars
acquire the majority of their mass is still unconfirmed.  If the ages of 
some Class I stars are indeed as old as T Tauri stars, the long-lived 
envelope lifetimes may stem from envelope replenishment, possibly caused 
by continued interactions with the cloud after formation.  Overall, it 
appears that most of Class I stars, as currently defined, have already
acquired the majority of their final stellar mass.

{\centering
\section{FUTURE PROSPECTS}
}

The ensemble of newly determined stellar and disk accretion properties of 
Class I stars offer powerful constraints on how and when young stars (and
brown dwarfs) are assembled.  However, many unknowns still remain.  Here 
we highlight 7 key areas of research that would help resolve the
remaining uncertainties and advance our understanding of the earliest
stages of star formation.

\noindent
{\it More Accurately Determined Circumstellar Properties -} Much of the
suspected misclassification of Class I stars could be confirmed or
refuted with more accurately determined SEDs based on observations over a
broad wavelength range which spatially resolve features (e.g., edge-on 
disks) and nearby neighbors.  In concert with this, more accurate and 
less orientation dependent criteria for identifying Class I stars needs 
to be established.

\noindent
{\it Extensive Surveys for Class I Stars -}
Larger, more complete (and less flux limited) surveys for Class I
stars in multiple star forming regions will help confirm or refute tentative
trends identified with the small samples studied so far, and may likewise
reveal real environmental (e.g., turbulence, gas temperature)
and/or (sub)stellar mass dependencies upon the formation process.

\noindent
{\it Improved Models of the Circumstellar Environment -}
With larger, more accurately determined samples of Class I stars, there 
will be a need for more sophisticated envelope-plus-disk models of 
embedded stars which can fit the observed SED and scattered 
light and polarization images (e.g., {\it Osorio et al.}, 2003; {\it Whitney  
et al.}, 2005; {\it Eisner et al.}, 2005; {\it Terebey et al.}, 2006).
This work is important for directly 
determining the density of the envelope material (which constrains the mass
infall rate), estimating the extinction to the central star (which is often
in error because of scattered light), and can potentially determine the 
system orientation.

\noindent
{\it Detailed Kinematic Mapping -} The case for massive infalling 
envelopes can be unambiguously resolved using interferometric 
techniques that kinematically map the surrounding envelope-like 
material.  In addition, these techniques offer the most direct 
and accurate way to determine the envelope infall rate, which can
be compared to the newly determined disk accretion rates.

\noindent
{\it Improved Models of Disk Accretion -}
Unfortunately there remain considerable uncertainties in observationally
determining disk accretion rates (e.g. bolometric correction), and
these uncertainties are magnified for embedded stars with 
high extinction and scattered light.  Consequently the absolute value 
of the disk accretion rates and their agreement with infall model 
predictions are difficult to assess.  Observations of emission-line 
profiles and continuum excess measurements over a broad range of wavelengths 
are promising methods to help resolve this.  
Additionally, observational monitoring to determine the timescale and
magnitude of variations in the mass accretion rate may yield important
constraints on how and how quickly mass is acquire during this stage.

\noindent
{\it More Sensitive Spectroscopic Surveys -} While the recent spectroscopic
observations focused upon here have revealed much about the pre-T Tauri 
evolutionary stage, the data suffer from rather severe observational biases.
These include biases in flux, extinction, and mass accretion rate.
More sensitive spectrographs and/or
larger aperture telescopes may be needed to address the first two biases,
while higher signal-to-noise observations of "featureless" Class I stars 
may help reveal their stellar and accretion properties.
Specifically, particular attention should be paid to the most luminous 
Class I stars 
in a given region, to establish better if they are more luminous because
they have accretion dominated luminosities, or if they are simply more
massive stars (e.g., L1551 IRS 5).

\noindent
{\it Comparisons with Synthetically Generated Spectra -}  
Fortunately in the last decade there has been considerable progress in the
area of synthetically generated spectra of stars.  The implication is
that stellar properties (e.g., temperature and surface gravity), can be
directly extracted from the spectra (e.g., {\it Johns-Krull et al.}, 1999, 
2004; {\it Doppmann et al.}, 2003, 2005), as opposed to indirectly
determined
by historic spectral-comparison techniques.  The most exciting application
is the determination of surface gravities, as the inferred stellar radii 
from these measurements can be used to establish more precise age
estimates, and perhaps unambiguously determine the age of Class I stars, 
even if only relative to T Tauri stars.

\bigskip
\textbf{Acknowledgments.} We thank Charles Lada for his substantial
contributions to many of the papers which constitute much of the work
discussed in this chapter. This work was partially supported by the
NASA Origins of Solar Systems program.  We have appreciated the
privilege to observe on the revered summit of Mauna Kea.

\bigskip

\centerline\textbf{ REFERENCES}
\bigskip
\parskip=0pt
{\small
\baselineskip=11pt

\refs Adams F. C.,  Lada C. J., and Shu F. J. (1987). {\it Astrophys. J.,
  312}, 788--806.

\refs Andr\'{e} P. and Montmerle T. (1994). {\it Astrophys. J., 420},
837--862. 

\refs Andr\'{e} P., Ward-Thompson D., and Barsony M. (1993). {\it Astrophys. J., 406}, 122--141.

\refs Andrews S. and Williams J. (2005). {\it Astrophys. J., 631}, 1134-1160.

\refs Armitage P. J. and Clarke C. J. (1996). {\it
  Mon. Not. R. Astron. Soc., 280}, 458--468. 

\refs Baraffe I., Chabrier G., Allard F., and Hauschildt P. H. (1998). {\it 
  Astron. Astrophys., 337}, 403--412. 

\refs Baraffe I., Chabrier G., Allard F., and Hauschildt P. H. (2002). {\it
  Astron. Astrophys., 382}, 563--572. 

\refs Bate M. R., Bonnell I. A., and Volker B. (2003). {\it
  Mon. Not. R. Astron. Soc., 339}, 577-599

\refs Beckwith S. V. W., Sargent A. I., Chini R. S., and G\"{u}sten
R. (1990). {\it Astrophys. J., 99}, 924--945. 

\refs Benson P. J. and Myers P. C. (1989). {\it Astrophys. J. Suppl., 71},
89--108. 

\refs Bontemps S., Andr\'{e} P., Kaas A. A., Nordh L., Olofsson G., et
al. (2001). {\it Astron. Astrophys., 372}, 173--194. 

\refs Boogert A. C. A., Blake G. A., and Tielens A. G. G. M. (2002). {\it
  Astrophys. J., 577}, 271--280.

\refs Bouvier J., Bertout C., Benz W., and Mayor M. (1986). {\it
  Astron. Astrophys., 165}, 110--119.

\refs Bouvier J., Cabrit S., Fern\'andez M., Martin E. L., and Matthews
J. M. (1993). {\it Astron. Astrophys., 272}, 176--206. 

\refs Bouvier J., Covino E., Kovo O., Martin E. L., Matthews J. M., et
al. (1995). {\it Astron. Astrophys., 299}, 89--107. 

\refs Brice\~{n}o C., Vivas A., Hern\'{a}ndez J., Calvet N., Hartmann L. et
al. (2004). {\it Astrophys. J., 606}, L123--126. 

\refs Brown D. W. and Chandler C. J. (1999). {\it
  Mon. Not. R. Astron. Soc., 303}, 855--863. 

\refs Calvet N. and Hartmann L. (1992). {\it Astrophys. J., 386}, 239--247.

\refs Calvet N., Hartmann L., and Strom S. E. (1997). {\it Astrophys. J.,
  481}, 912--917. 

\refs Calvet N., Hartmann L., and Strom S. E. (2000). In {\it Protostars
  and Planets IV} (V. Mannings et al., eds.), pp. 377--399. Univ. of
Arizona, Tucson.  

\refs Casali M. M. and Eiroa C. (1996). {\it Astron. Astrophys., 306},
427--435. 

\refs Casali M. M. and Matthews H. E. (1992). {\it
  Mon. Not. R. Astron. Soc., 258}, 399--403. 

\refs Cassen P. and Moosman A. (1981). {\it Icarus, 48}, 353--376.

\refs Chiang E. I. and Goldreich P. (1999). {\it Astrophys. J., 519},
279--284. 

\refs Cieza, L. A., Kessler-Silacci, J. E., Jaffe, D. T., Harvey P. M., and
Evans N. J. II (2005). {\it Astrophys. J., 635}, 422--441.

\refs Clarke C. J. and Bouvier J. (2000). {\it Mon. Not. R. Astron. Soc.,
  319}, 457--466. 

\refs Cohen M. and Schwartz R. D. (1983). {\it Astrophys. J., 265},
877--900. 

\refs Collier Cameron A. and Campbell C. G. (1993). {\it
  Astron. Astrophys., 274}, 309--318. 

\refs Covey K. R., Greene T. P., Doppmann G. W., and Lada
C. J. (2005). {\it Astron. J., 129}, 2765--2776. 

\refs D'Antona F. and Mazzitelli I. (1997).  {\it Mem. Soc.
  Astron. Ital., 68}, 807--822. 

\refs Doppmann G. W., Jaffe D. T., and White R. J.  (2003). {\it
  Astron. J., 126}, 3043--3057. 

\refs Doppmann G. W., Greene T. P., Covey K. R., and Lada
C. J. (2005). {\it Astron. J., 130}, 1145--1170. 

\refs Duch\^{e}ne G., Bouvier J., Bontemps S., Andr\'e P., and Motte
 F. (2004). {\it Astron. Astrophys., 427}, 651--665.

\refs Edwards S., Strom S., Hartigan P., Strom K., Hillenbrand L., et al.
(1993). {\it Astron. J., 106}, 372--382. 

\refs Eisner J. A., Hillenbrand L. A., Carpenter J. M., and Wolf
S. (2005). {\it Astrophys. J., 635}, 396--421.

\refs Evans N. J. (1999). {\it Ann. Rev. Astron. Astrophys., 37}, 311-362.

\refs Folha D. F. M. and Emerson J. P. (1999). {\it Astron. Astrophys.,
  352}, 517--531. 

\refs Goodwin S. P., Whitworth A. P., and Ward-Thompson D. (2004). {\it
  Astron. Astrophys., 423}, 169--182

\refs Graham J. A. (1991). {\it Publ. Astron. Soc. Pac., 103}, 79--84.

\refs Greene T. P. and Lada C. J. (1996). {\it Astron. J., 112}, 2184--2221.

\refs Greene T. P. and Lada C. J. (1997). {\it Astron. J., 114}, 2157--2165.

\refs Greene T. P. and Lada C. J. (2000). {\it Astron. J., 120}, 430--436.

\refs Greene T. P. and Lada C. J. (2002). {\it Astron. J., 124}, 2185--2193.

\refs Gregersen E. M., Evans N. J., Zhou S., and Choi M. (1997). {\it
  Astrophys. J., 484}, 256--276. 

\refs Gullbring E., Hartmann L., Brice\~{n}o C., Calvet N. (1998). {\it
  Astrophys. J., 492}, 323--341. 

\refs Haisch K. E. Jr., Greene T. P., Barsony M., and Stahler
S. W. (2004). {\it Astrophys. J., 127} 1747--1754.

\refs Hartigan P., Edwards S., and Ghandour L. (1995). {\it Astrophys. J.,
  452}, 736-768. 

\refs Hartmann L. (1998). {\it Accretion Processes in Star Formation},
(Cambridge: Cambridge Univ. Press), chap. 4.

\refs Hartmann L., and Kenyon S. J. (1987). {\it Astrophys. J., 312},
243--253. 

\refs Hartmann L., and Kenyon S. J. (1996). {\it
  Ann. Rev. Astron. Astrophys., 34}, 207--240. 

\refs Hartmann L., Hewett R., Stahler S., and Mathieu R. D. (1986). {\it
  Astrophys. J., 309}, 275--293. 

\refs Hartmann L., Hewett R. and Calvet N. (1994). {\it Astrophys. J.,
  426}, 669--687. 

\refs Henning T., Begemann B., Mutschke H., and Dorshner J. (1995). {\it
  Astron. Astrophys. Suppl., 112}, 143

\refs Hillenbrand L. A., Strom S. E., Calvet N., Merrill K. M., Gatley,
I. et al. (1998). {\it Astron. J., 116}, 1816--1841.

\refs Hogerheijde M. R. and Sandell G. (2000). {\it Astrophys. J., 534},
880--893. 

\refs Ishii M., Tamura M., and Itoh, Y. (2004). {\it Astrophys. J., 612}, 
956--965 

\refs Johns-Krull C. M., Valenti J. A., and Koresko C. (1999). {\it
  Astrophys. J., 516}, 900--915. 

\refs Johns-Krull C. M., Valenti J. A., and Gafford A. D. (2003). {\it
  Rev. Mex. Astron. Astrofys., 18}, 38--44. 

\refs Johns-Krull C. M., Valenti J. A., and Saar S. H. (2004). {\it
  Astrophys. J., 617}, 1204--1215. 

\refs Kaas A. A., Olofsson G., Bontemps S., Andr\'e P., Nordh L., et
al. (2004). {\it Astron. Astrophys., 421}, 623--642.

\refs Kenyon S. J., Calvet N., and Hartmann L. (1993a). {\it Astrophys. J.,
  414}, 676--694.  

\refs Kenyon S. J., Whitney B. A., G\'omez M., and Hartmann L. (1993b). {\it
  Astrophys. J., 414}, 773-792.  

\refs Kenyon S. J., G\'omez M., Marzke R. O., and Hartmann L. (1994). {\it
  Astron. J., 108}, 251--261. 

\refs Kenyon S. J. and  Hartmann L. (1995). {\it Astrophys. J. Suppl.,
  101}, 117--171. 

\refs Kenyon S. J., Hartmann L. W., Strom K. M., and Strom
S. E. (1990). {\it Astron. J., 99}, 869--887. 

\refs Kenyon S. J., Brown D. I., Tout C. A., and Berlind P. (1998). {\it
  Astron. J., 115}, 2491--2503. 

\refs Kessler-Silacci J. E., Hillenbrand L. A., Blake G. A., and Meyer
 M. R. (2005). {\it Astrophys. J.}, 622, 404--429

\refs Kleinmann S. G. and Hall D. N. B. (1986). {\it Astrophys. J. Suppl.,
  62}, 501--517. 

\refs K\"onigl A. (1991). {\it Astrophys. J., 370}, L39--43.

\refs Lada C. J. (1987). In {\it IAU Symp.~115: Star Forming Regions}
(M. Peimbert and J. Jugaku, eds.), pp. 1--18., D. Reidel Publishing Co., 
Dordrecht.

\refs Lada C. J. and Wilking, B. A (1984). {\it Astrophys. J., 287},
610--621. 

\refs Ladd E. F., Lada E. A., and Myers P. C. (1993). {\it Astrophys. J.,
  410}, 168--178. 

\refs Larson R. B. (1984). {\it Mon. Not. R. Astron. Soc., 206}, 197--207.

\refs Luhman K. L. and Rieke G. H. (1999). {\it Astrophys. J., 525},
440--465. 

\refs Mac Low M. and Klessen R. S. (2004). {\it Rev. Mod. Phys., 76},
125--194. 

\refs Matt S. and Pudritz R. (2005). {\it Astrophys. J., 632}, L135--138. 

\refs Motte F. and Andr\'e P. (2001). {\it Astron. Astrophys., 365},
440--464. 

\refs Mundt R., Stocke J., Strom S. E., Strom K. M., and Anderson
E. R. (1985). {\it Astrophys. J., 297}, L41--45. 

\refs Muzerolle J., Calvet N., and Hartmann L. (1998). {\it Astrophys. J.,
  492}, 743--753. 

\refs Muzerolle J., Hartmann L., and Calvet N. (1998). {\it Astron. J.,
  116}, 455--468. 

\refs Muzerolle J., Hartmann L., and Calvet N. (1998). {\it Astron. J.,
  116}, 2965--2974. 

\refs Muzerolle J., Calvet N., and Hartmann L. (2001). {\it Astrophys. J.,
  550}, 944-961. 

\refs Muzerolle J., D'Alessio P., Calvet N., and Hartmann L. (2004). {\it
  Astrophys. J., 617}, 406. 

\refs Myers P. C. and Benson P. (1983). {\it Astrophys. J., 266},
309--320. 

\refs Myers P. C. and Ladd E. F. (1993). {\it Astrophys. J., 413},
L47--50. 

\refs Myers P. C., Fuller G. A., Mathieu R. D., Beichman C. A., Benson
P. J., et al. (1987). {\it Astrophys. J., 319},
340--357.

\refs Najita J. (2004). in {\it Star Formation in the Interstellar Medium:
  In Honor of David Hollenbach, Chris McKee and Frank Shu} (Johnstone D.,
  Adams F. C., Lin D. N. C., Neufeld D. A., and Ostriker E. C., eds.)
  pp. 271--277. ASP, Provo. 

\refs Nisini B., Antoniucci S., Giannini T., and Lorenzetti D. (2005). {\it
  Astron. Astrophys., 429}, 543--557. 

\refs Onishi R., Mizuno A., Kawamura A., Tachihara K., and Fukui
Y. (2002). {\it Astrophys. J, 575}, 950--973 

\refs Osorio M., D'Alessio P., Muzerolle J., Calvet N., and Hartmann
L. (2003). {/it Astrophys. J., 586}, 1148--1161. 

\refs Osterloh M. and Beckwith S. V. W. (1995). {\it Astrophys. J., 439},
288--302. 

\refs Padgett D. L., Brandner W., Stapelfeldt K. R., Strom S. E., Terebey
S., et al. (1999). {\it Astron. J., 117}, 1490--1504. 

\refs Park S. and Kenyon S. J. (2002). {\it Astron. J., 123}, 3370--3379.

\refs Prusti T., Whittet D. C. B., and Wesselius P. R. (1992). {\it
  Mon. Not. R. Astron. Soc., 254}, 361--368. 

\refs Rebull L. M., Wolff S. C., Strom S. E., and Makidon
R. B. (2002). {\it Astron. J., 124}, 546--559. 

\refs Rebull L. M., Wolff S. C., and Strom S. E. (2004). {\it Astron. J.,
  127}, 1029--1051. 

\refs Rhode K. L., Herbst W., and Mathieu R. D (2001). {\it Astron. J.,
  122}, 3258--3279. 

\refs Safier P. N. (1998). {\it Astrophys. J., 494}, 336--341.

\refs Shu F. H. (1977). {\it Astrophys. J., 214}, 488--497.

\refs Shu F., Najita J., Ostriker E., Wilkin F., Ruden S., et al.
(1994). {\it Astrophys. J., 429}, 781--796. 

\refs Siess L., Forestini M., and Bertout C. (1999). {\it
  Astron. Astrophys., 342}, 480--491. 

\refs Sonnhalter C., Preibisch T., and Yorke H. W. (1995). {\it
  Astron. Astrophys., 299}, 545--556. 

\refs Stassun K. G., Mathieu R. D., Mazeh T., and Vrba F. (1999). {\it
  Astron. J., 117}, 2941--2979. 

\refs Stassun K., Mathieu R., Vrba J., Mazeh T., and Henden A. (2001). {\it
  Astron. J., 121}, 1003--1012. 

\refs Terebey S., Shu F. H., and Cassen P. (1984). {\it Astrophys. J.,
  286}, 529--551. 

\refs Terebey S., Van Buren D., and Hancock T. (2006). {\it Astrophys. J.,
  637}, 811--822

\refs Tout C. A., Livio M., and Bonnell I. A. (1999). {\it
  Mon. Not. R. Astron. Soc., 310}, 360--376. 

\refs Wallace L. and Hinkle K. (1996). {\it Astrophys. J. Suppl., 107},
312--390. 

\refs Werner, M. W., Roellig T. L., Low F. J., Rieke G. H., Rieke M., et 
al. (2004). {\it Astrophys. J. Suppl., 154}, 154--162.

\refs White R. J. and Basri G. (2003). {\it Astrophys. J., 582},
1109--1122. 

\refs White R. J. and Hillenbrand L. A. (2004). {\it Astrophys. J., 616},
998--1032. 

\refs Whitney B. A., Kenyon S. J., and G\'omez M. (1997). {\it Astrophys. J.,
  485}, 703--734. 

\refs Whitney B. A., Wood K., Bjorkman J. E., and Cohen M. (2003). {\it
  Astrophys. J., 598}, 1079--1099. 

\refs Whitney B. A., Indeebetouw R., Bjorkman J. E., and Wood K. (2004). 
 {\it Astrophys. J., 617}, 1177-1190.

\refs Whitney B. A., Robitaille T. P., Wood K., Denzmore P. and Bjorkman
J. E. (2005) in {\it PPV Poster Proceedings} \\ 
http://www.lpi.usra.edu/meetings/ppv2005/pdf/8460.pdf

\refs Wilking B. A., Lada C. J., and Young E. T. (1989). {\it
  Astrophys. J., 340}, 823--852. 

\refs Wilking b. A., Greene T. P., Lada C. J., Meyer M. R., and Young
E. T. (1992). {\it Astrophys. J., 397}, 520-533 

\refs Wolf S., Padgett D. L., and Stapelfeldt K. R. (2003). {\it
  Astrophys. J., 588}, 373--386. 

\refs Yorke H. W., Bodenheimer P., and Laughlin G. (1993). {\it
  Astrophys. J., 411}, 274--284. 

\refs Young C. H., Shirley Y. L., Evans N. J., and Rawlings
J. M. C. (2003). {\it Astrophys. J. Suppl., 145}, 111--145. 

\end{document}